\documentclass[10pt]{iopart}
\usepackage{epsf,epsfig}

\newcommand{\kb}{\bar k}
\renewcommand{\d}{{\rm d}}
\newcommand{\bk}{{\bf k}}
\newcommand{\bx}{{\bf x}}
\newcommand{\bq}{{\bf q}}

\begin{document}

\title{Eigenmodes of 3-dimensional spherical spaces
       and their application to cosmology}

\author{Roland Lehoucq$^{1,6}$,
        Jeffrey Weeks$^{2}$,
        Jean-Philippe Uzan$^{3,4}$
        Evelise Gausmann$^{5}$, and
        Jean-Pierre Luminet$^{6}$
        }

\date{DRAFT VERSION 2: 12 february 2002}

\address{(1) CE-Saclay, DSM/DAPNIA/Service d'Astrophysique,
             F-91191 Gif sur Yvette Cedex (France)\\
         (2) 15   Farmer   St.,  Canton  NY  13617-1120  (USA).\\
         (3) Institut d'Astrophysique de Paris, GReCO, CNRS-FRE 2435,\\
             98 bis, Bd Arago, 75014 Paris
             (France).\\
         (4) Laboratoire de Physique Th\'eorique, CNRS-UMR 8627,
             B\^at. 210, Universit\'e Paris XI, F--91405 Orsay Cedex
(France),\\
         (5) Instituto de F\'{\i}sica Te\'orica, Rua Pamplona, 145 Bela Vista
            - S\~ao Paulo - SP, CEP 01405-900 (Brasil)\\
         (6) Laboratoire Univers et Th\'eories, CNRS-FRE 2462,
             Observatoire  de  Paris, F-92195 Meudon Cedex (France).
        }

\begin{abstract}
   This article investigates the computation of the eigenmodes of the
   La\-pla\-cian operator in multi-connected three-dimensional
   spherical spaces.  General mathematical results and analytical
   solutions for lens and prism spaces are presented.  Three
   complementary numerical methods are developed and compared with our
   analytic results and previous investigations.  The cosmological
   ap\-pli\-ca\-tions of these results are discussed, focusing on the
   cosmic microwave background  (CMB) anisotropies. In particular,
   whereas in the  Euclidean case too small universes are excluded by
   present CMB data, in the spherical case there will always exist
   candidate topologies even if the total energy density parameter of
   the universe is very close to unity.

\end{abstract}


\section{Introduction}

The search for the topology of our universe has made tremendous
progress in the past years and several methods have been designed
using either galaxy catalogs or the CMB (see
e.g.~\cite{lachieze95,uzan99} for general reviews).  The most
promising dataset that can eventually contain a topological signature
is the cosmic microwave background (CMB) in the form of pattern
correlations (such as homologous circles in the sky~\cite{cornish98},
or anomalously large temperature correlations in a set of
directions~\cite{levin98}, see~\cite{levin02} for a recent review on
the CMB methods) or non Gaussianity~\cite{inoue00}.

The detectability of the topology in datasets such as those that will
be made available by the MAP~\cite{map} and Planck~\cite{planck}
satellite missions requires to simulate maps with the topological
signature for a large set of topologies.  These maps will have mainly
two uses: first, they will allow us to test the detection method and
for instance estimate its running time and second, once all sources of
noises are added, it will help us investigating to which extent a
given method is well suited to detect the topological signal and
indeed if it is not blurred (in the same spirit as the investigation
of the ``crystallographic" methods based on galaxy
catalogs~\cite{us3}).  A prerequisite for any further study is thus to
simulate CMB maps with a topological signal.

At present, the status of the constraint on the topology of the
universe is sparse.  Concerning locally Euclidean spaces, it was shown
on the basis of the COBE data that the size of the fundamental domain
of a 3-torus has to be larger than $L\geq4800\,h^{-1}$~Mpc
\cite{sokolov93,staro93,stevens93,costa95}.  This constraint does not
exclude a toroidal universe since there can be up to $N=8$ copies of
the fundamental cell within our horizon.  This constraint relies
mainly on the fact that the smallest wavenumber is $2\pi/L$, which
induces a suppression of fluctuations on scales beyond the size of the
fundamental domain.  This result holds only for the case of a
vanishing cosmological constant and was generalized to all Euclidean
manifolds~\cite{levin99}.  A non-vanishing cosmological constant
induces larger scale cosmological perturbations, via the integrated
Sachs-Wolfe effect.  For instance if $\Omega_\Lambda=0.9$ and
$\Omega_m=0.1$, the former is relaxed to allow for $N=49$ copies of
the fundamental cell within our horizon.  This constraint is also
milder in the case of compact hyperbolic manifolds and it was
shown~\cite{aurich00,inoue00b,cornish00} that the angular power
spectrum was consistent with the COBE data on multipoles ranging from
2 to 20 for the Weeks and Thurston manifolds.  Another approach was
developed in~\cite{bond1,bond2,bond3} and is based on the method of
images.  Only one spherical space using this method of images was
considered in the literature, namely the case of the projective
space~\cite{souradeep}.

Note that multiconnectedness breaks global homogeneity and isotropy
(except for the particular case of the projective space).  It follows
that the temperature angular correlation function $C$ will depend on
the position of the observer and on the orientation of the manifold,
which is at odd with the standard lore.  In a simply connected
universe the angular correlation function depends only on the angle
between the two directions whereas in a multi-connected universe, it
will depend on the two directions.  It follows that the coefficients
$C_\ell$ of the decomposition of $C$ into Legendre polynomials,
obtained by averaging over the sky, loose much of topological
information.  As clearly explained in~\cite{levin02}, $C$ can be
decomposed into an isotropic and an anisotropic part and the $C_\ell$
depend solely on the former.  The $C_\ell$ alone are a poor indicator
of the topology, despite the fact that they can help constraining the
topology, which backs up the necessity to study the full sky map.

In standard relativistic cosmology, the universe is described by a
Friedmann-Lema\^{\i}tre spacetime with locally isotropic and
homogeneous spatial sections.  These spatial sections can be defined
as the constant density or time hypersurfaces.  With such a
splitting,
the equations of evolution of the matter and geometry perturbations
that will give birth to the large scale structures of the universe
reduce to a set of coupled differential equations involving a
Laplacian (see e.g.~\cite{kodama}).  This system is conveniently
(numerically) solved in Fourier space but this requires to determine
the eigenmodes and eigenvalues of the Laplacian through the
generalized Helmoltz equation
\begin{equation}
\label{Helmoltz1}
\Delta \Psi_{q} = -q^2\Psi_{q}.
\end{equation}
The Laplacian in Eq.~(\ref{Helmoltz1}) is defined as $\Delta\equiv
D_iD^i$, $D_i$ being the covariant derivative associated with the
metric $\gamma_{ij}$ of the spatial sections ($i,j=1..3$). The
eigenmodes on which any function can be developed encode the boundary
conditions imposed by the topology, and any function developed on
this
basis will satisfy the required boundary conditions.

Concerning the computation of the eigenmodes, the case of Euclidean
manifolds can be solved analytically and many numerical investigations
of compact hyperbolic manifolds have been
performed~\cite{inoue00b,aurich89,inoue99,aurich96,cornish99}.  The
case of spherical manifolds has been disgarded during a long time.
The recent measurements of the density parameters let the room for our
universe to be slightly positively curved since they are estimated to
lie in the range $\Omega_{0} \equiv
\Omega_{m_{0}}+\Omega_{\Lambda_{0}} = 1.11^{+0.13}_{-0.12}$ to $95 \%$
confidence~\cite{jaffe00}.

The goal of this article is to focus on the computation of the
eigenmodes of the Laplacian in spherical spaces, having in mind their
use to simulate CMB maps.  In Section~\ref{sec_math}, we recall the
main mathematical results on the classification of spherical
spaces~\cite{glluw} and some analytical results on the eigenmodes of
the Laplacian such as the determination of the smallest wavenumber and
its multiplicity.  In section~\ref{sec_num}, we describe the three
numerical methods that are used to compute the eigenvalues and
eigenmodes and we discuss their precision and efficiency.  We then
describe briefly, in Section~\ref{sec_torus}, a method to obtain these
eigenfunctions analytically in the particular cases of lens and prism
spaces.  Such a result is of importance to compare with the output of
numerical computations.  After describing some statistical properties
of the eigenmodes in Section~\ref{sec_stat}, we present, in
Section~\ref{sec_cosmo}, some cosmological consequences of our result.
We estimate the effect of the topology on the Sachs-Wolfe plateau.
\ref{appA} gathers most of the results on the eigenfunctions of the
Laplacian in homogeneous and isotropic 3-dimensional simply connected
spaces.  \vskip0.5cm

\noindent{\bf Notations:} The local geometry of the universe is
described
by a Friedmann--Lema\^{\i}tre metric
\begin{equation}\label{fl_metric}
\d s^2=-\d t^2+a^2(t)\left[\d\chi^2+f^2({\chi})\d\Omega^2
\right]
\end{equation}
with $f(\chi)=({\rm sinh}\chi,\chi,\sin\chi)$ respectively for
hyperbolic, Euclidean and spherical spatial sections. $a(t)$ is
the scale factor, $t$ the cosmic time and $\d\Omega^2\equiv
\d\theta^2+\sin^2{\theta}\d\varphi^2$ the infinitesimal solid
angle. $\chi$ is the (dimensionless) comoving radial distance in
units of the curvature radius $R_C$.

We use the embedding of the 3-sphere $S^3$ in 4-dimensional
Euclidean space by introducing the set of coordinates
$(x_\mu)_{\mu=0..3}$ related to the intrinsic comoving coordinates
$(\chi,\theta,\varphi)$ through (see e.g. Ref.~\cite{wolfe67})
\begin{eqnarray}\label{klein}
x_0&=&\cos{\chi}\nonumber\\
x_1&=&\sin{\chi}\sin{\theta}\sin{\varphi}\nonumber\\
x_2&=&\sin{\chi}\sin{\theta}\cos{\varphi} \nonumber\\
x_3&=&\sin{\chi}\cos{\theta},
\end{eqnarray}
with $0\leq\chi\leq\pi$, $0\leq\theta\leq\pi$ and
$0\leq\varphi\leq2\pi$.
The 3--sphere is then the submanifold of equation
\begin{equation}
x^\mu x_\mu \equiv x_0^2+x_1^2+x_2^2+x_3^2=+1,
\end{equation}
where $x_\mu=\delta_{\mu\nu}x^\nu$.  The comoving spatial distance $d$
between any two points $x$ and $y$ on $S^3$ can be computed using the
inner product $x^\mu y_\mu$ and is given by
\begin{equation}
d[x,y]={\rm arc}\cos{\left[x^\mu y_\mu\right]}.
\end{equation}
The volume enclosed by a sphere of radius $\chi$ is, in units of the
curvature radius,
\begin{equation}\label{vol_chi}
\mathrm{Vol}(\chi)=\pi\left(2\chi-\sin{2\chi}\right).
\end{equation}

We consider 3-dimensional multi-connected manifolds of the form
${\cal M}_3=S^3/\Gamma$ where the holonomy group\footnote{We use
this term to mean the geometric version of the group of covering
transformations, that is, elements of the holonomy group are
isometries, while covering transformations are homeomorphisms,
see~\cite{thurston97}.} $\Gamma$ is a discrete subgroup of $SO(4)$
that acts without fixed point on the 3-dimensional covering space
$S^3$. It is isomorphic to the first fondamental group
$\pi_1({\cal M}_3)$. The elements $g\in\Gamma$ are isometries and
can be expressed as $4\times4$ matrices acting on the elements of
the 4-dimensional embedding Euclidean space.

The transformation (\ref{klein}) between the intrinsic coordinates
$(\chi,\theta,\phi)$ and $x^\mu$ is bijective. In the following of
the article, if $f$ is a function on $S^3$, we will loosely write
$f(x)$ for $f(\chi,\theta,\phi)$ and for instance $f(gx)$ will
refer to $f(\chi',\theta',\phi')$ with $(\chi',\theta',\phi')$
being the intrinsic coordinates of the image $x'=gx$ obtained from
Eq.~(\ref{klein}) [see e.g. Eq.~(\ref{A14}) for an example].

To finish, let us remark that any eigenmode of a multi-connected
manifold $S^3/\Gamma$ lifts to a $\Gamma$-invariant\footnote{A
$\Gamma$-invariant function $f$ is a function satisfying
$f(x)=f(gx)$ for all $g\in\Gamma$ and for all $x$.} eigenmode of
$S^3$, and conversely each $\Gamma$-invariant eigenmode of $S^3$
projects down to an eigenmode of $S^3/\Gamma$. Thus, with a slight
abuse of terminology, we may say that the eigenmodes of
$S^3/\Gamma$ are the $\Gamma$-invariant eigenmodes of $S^3$.

\section{Mathematical results on the Laplacian operator in spherical
spaces}\label{sec_math}

\subsection{Classification of spherical manifolds}

In our preceding article~\cite{glluw}, we presented in a pedestrian
way the complete classification of 3-dimensional spherical topologies
and we described how to compute their holonomy transformations.  We
briefly outline the main results of this study, mainly to set our
notations.  The classification of constant curvature spherical
manifolds was first presented in~\cite{threlfall,wolfe67}.

The isometry group of the 3-sphere is $G=SO(4)$ and one needs to
determine all the finite subgroups of $SO(4)$ acting fixed point
freely.  Every isometry of $SO(4)$ can be uniquely decomposed as the
product of a right-handed ($R$) and a left-handed ($L$) Clifford
translation\footnote{Clifford translations are isometries that
translate all points the same distance.}, up to a multiplication by
-1
of both factors.  Besides, $S^3$ enjoys a group structure as the set
${\cal S}^3$ of unit length quaternions.  It can then be shown that
each right-handed (resp.  left-handed) Clifford translation
corresponds to a left (resp.  right) multiplication of ${\cal S}^3$
[${\bf q}\rightarrow {\bf x}{\bf q}$ (resp.  ${\bf q}\rightarrow {\bf
q}{\bf x}$)] so that the two groups of right-handed and left-handed
Clifford translations are isomorphic to ${\cal S}^3$.  It follows
that
$SO(4)$ is isomorphic to ${\cal S}^3\times{\cal
S}^3/\lbrace\pm({\bf1},{\bf1})\rbrace$ so that the classification of
the subgroups of $SO(4)$ can be deduced from the classification of
all
subgroups of ${\cal S}^3$.  It can then be shown that there exists a
two-to-one homomorphism from ${\cal S}^3$ to $SO(3)$ the subgroups of
which are known.  It follows that the finite subgroups of ${\cal
S}^3$
are:
\begin{itemize}
\item   the cyclic groups $Z_n$ of order $n$,
\item   the binary dihedral groups $D_m^*$ of order $4m$, $m \ge 2$,
\item   the binary tetrahedral group $T^*$ of order 24,
\item   the binary octahedral group $O^*$ of order 48,
\item   the binary icosahedral group $I^*$ of order 120,
\end{itemize}
where a binary group is the two-fold cover of the corresponding group.

>From this classification, it can be shown that there are three
categories of spherical 3-manifolds.
\begin{itemize}
\item The {\it single action manifolds} are those for which a
subgroup $R$
of ${\cal S}^3$ acts as pure right-handed Clifford translations. They
are thus the simplest spherical manifolds and can all be written as
$S^3/\Gamma$
with $\Gamma=Z_n,D_m^*,T^*,O^*,I^*$.
\item The {\it double-action manifolds} are those for which subgroups
$R$ and $L$ of ${\cal S}^3$ act simultaneously as right- and
left-handed Clifford translations, and every element of $R$ occurs
with every element of $L$. There are obtained for the groups
$\Gamma=\Gamma_1\times \Gamma_2$ with
$(\Gamma_1,\Gamma_2)=(Z_m,Z_n), (D^*_m,Z_n), (T^*,Z_n), (O^*,Z_n),
(I^*,Z_n)$ respectively with ${\rm gcd}(m,n)=1$, ${\rm
gcd}(4m,n)=1$, ${\rm gcd}(24,n)=1$, ${\rm gcd}(48,n)=1$, ${\rm
gcd}(120,n)=1$.
\item The {\it linked-action manifolds} are similar to the double
action
manifolds, except that each element of $R$ occurs with only some of
the
elements of $L$.
\end{itemize}
The classification of these manifolds is summarized in the
figure~8 of~\cite{glluw}.

We also define a lens space $L(p,q)$ by identifying the lower surface
of
a lens-shaped solid to the upper surface with a ${2\pi}q/p$ rotation
for
relatively prime integers $p$ and $q$ with $0 < q < p$.  Furthermore,
we
may restrict our attention to $0 <q \leq p/2$ because for values of
$q$
in the range $p/ 2 < q < p$ the twist ${2\pi}q/p$ is the same as
$-{2\pi}(p-q)/p$, thus $L(p,q)$ is the mirror image of $L(p,p-q)$.
Lens
spaces can be of any of the category described above and their
classification is detailed in figure~9 of~\cite{glluw}.

\subsection{Spectrum of spherical manifolds}

The Helmoltz equation~(\ref{Helmoltz1}) is usually rewritten in terms
of the quantity $\beta$ defined as
\begin{equation}
\beta^2\equiv q^2+1
\end{equation}
and takes the form
\begin{equation}
\Delta\Psi_{k,s}=-k(k+2)\Psi_{k,s},
\end{equation}
after using the change of variable $\beta = k+1$ and where $s$ is an
integer labelling the modes of same eigenvalue (see
Ref.~\cite{helgason} for the properties of the Laplacian operator).
It follows that the eigenvalues of the Laplacian on $S^3$ are
$k(k+2)$, $k$ being an integer, and that the multiplicity of each
eigenvalue is $(k+1)^2$ [see~\ref{appA} for details].  The two modes
$k=0$ and $k=1$ are gauge modes since they respectively correspond to
a change in the curvature radius (homogeneous deformation) and in a
displacement of the center of the 3-sphere and are thus physically not
relevant \cite{Lifshitz}.  It is thus clear that the eigenvalues of
the Laplacian on a multi-connected spherical manifold ${\cal M}_3=S^3/
\Gamma$ will be a subset of the eigenvalues $k(k+2)$ of the Laplacian
on $S^3$.

For any spherical manifold ${\cal M}_3$, we can introduce its
(discrete) spectrum as the set of all eigenvalues of the Laplacian
\begin{equation}\label{ikeda1}
{\rm Sp}({\cal M}_3)=\lbrace
0=\lambda_0<\lambda_1\leq\lambda_2\leq\ldots
\leq\lambda_i\leq\ldots\rbrace
\end{equation}
and two Riemannian manifolds ${\cal M}_3$ and ${\cal M'}_{3}$ are
said to be isospectral if ${\rm Sp}({\cal M}_3)={\rm Sp}({\cal
M'}_3)$.

Ikeda and Yamamoto~\cite{Ikeda79} studied the spectra of 3-dimensional
lens spaces and demonstrated that if two 3-dimensional lens spaces
with fundamental group of order $q$ are isospectral to each other,
then they are isometric to each other.  Ikeda~\cite{Ikeda80a} extended
this result to show that if two 3-dimensional spherical manifolds are
isospectral then they are isometric.  Combining this with previous
results, it follows that a 3-dimensional spherical manifold is
completely
determined, as a Riemannian manifold, by its spectrum.
Higher-dimensional generalizations appear in~\cite{Ikeda80b}.

\begin{figure}
\begin{center}
\epsfig{file=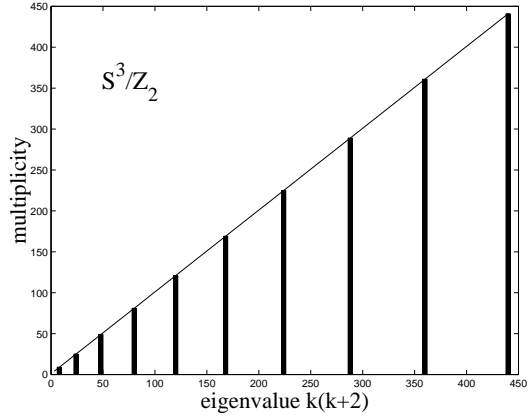,width=7cm}
\caption{Comparison of the spectra of the 3-sphere and of the
projective space. Half of the modes are lost due to the reflection
symmetry (since antipodal points are identified, the eigenfunction
must vanish on the equator).} \label{fig1}
\end{center}
\end{figure}

\begin{figure}
\begin{center}
\epsfig{file=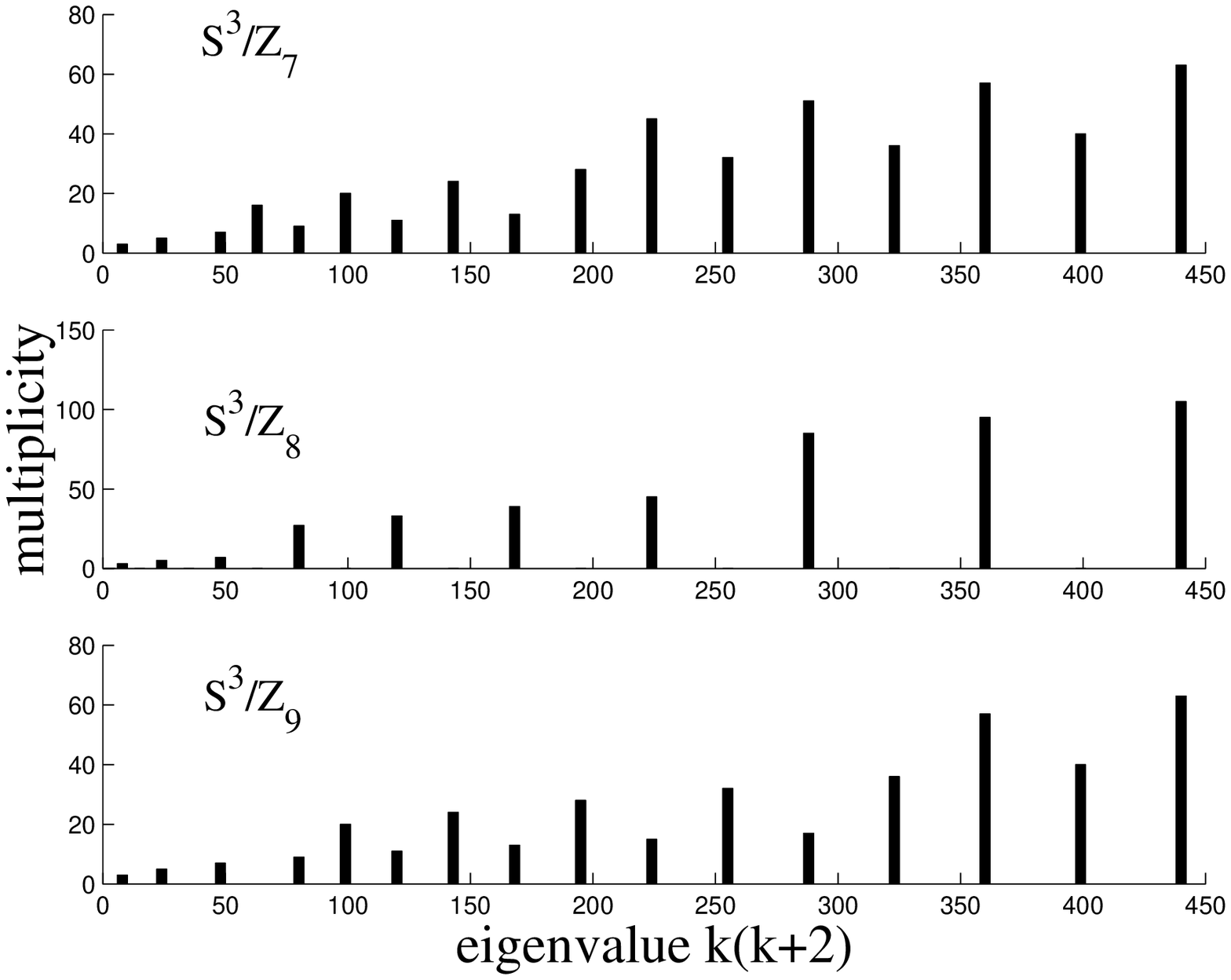,width=6cm}
\epsfig{file=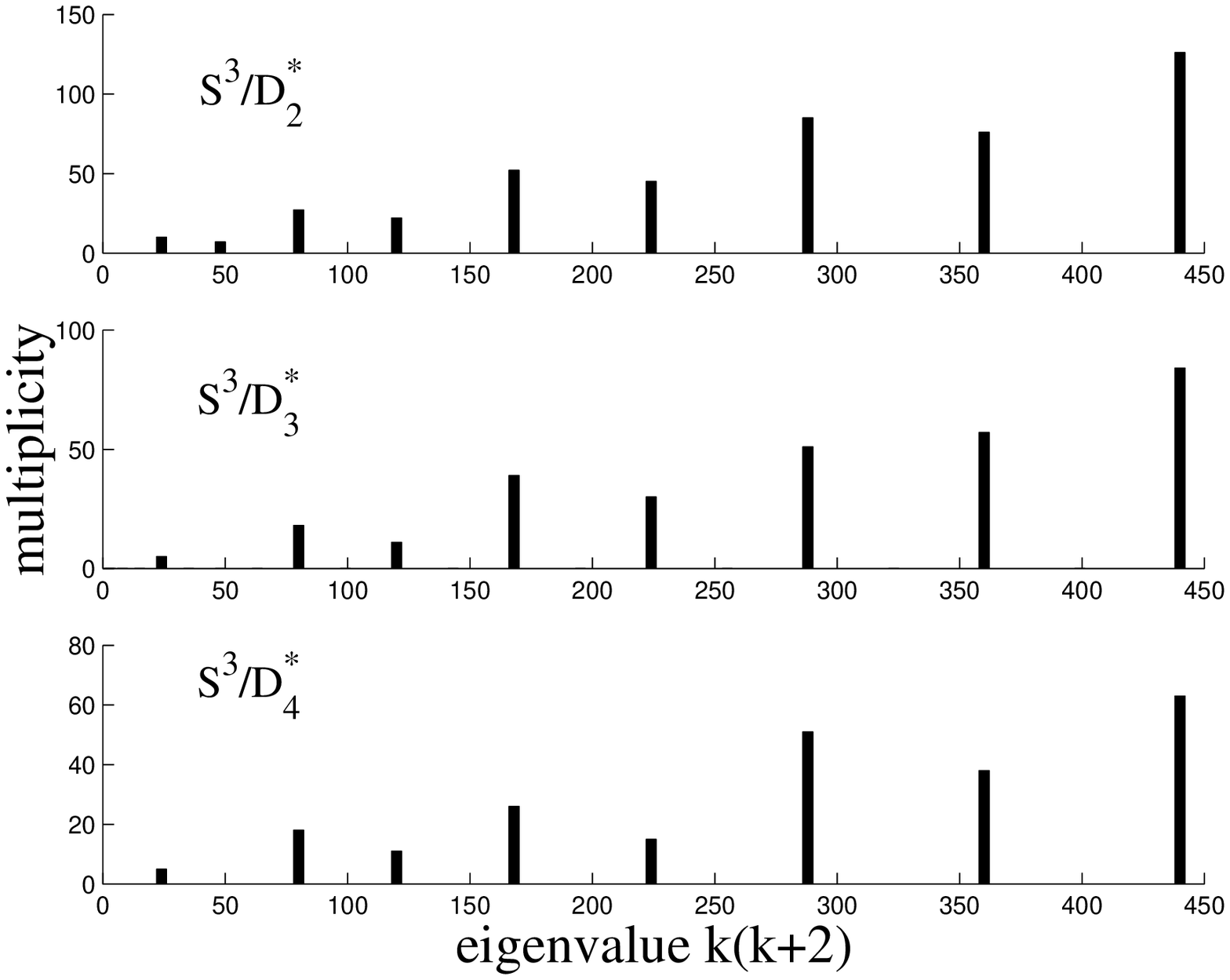,width=6cm} \caption{Spectra
of single-action manifolds generated from [left] the cyclic groups
$Z_7$, $Z_8$ and $Z_9$ and from [right] the binary dihedral groups
$D_2^*$, $D_3^*$ and $D_4^*$.} \label{fig2}
\end{center}
\end{figure}

\begin{figure}
\begin{center}
\epsfig{file=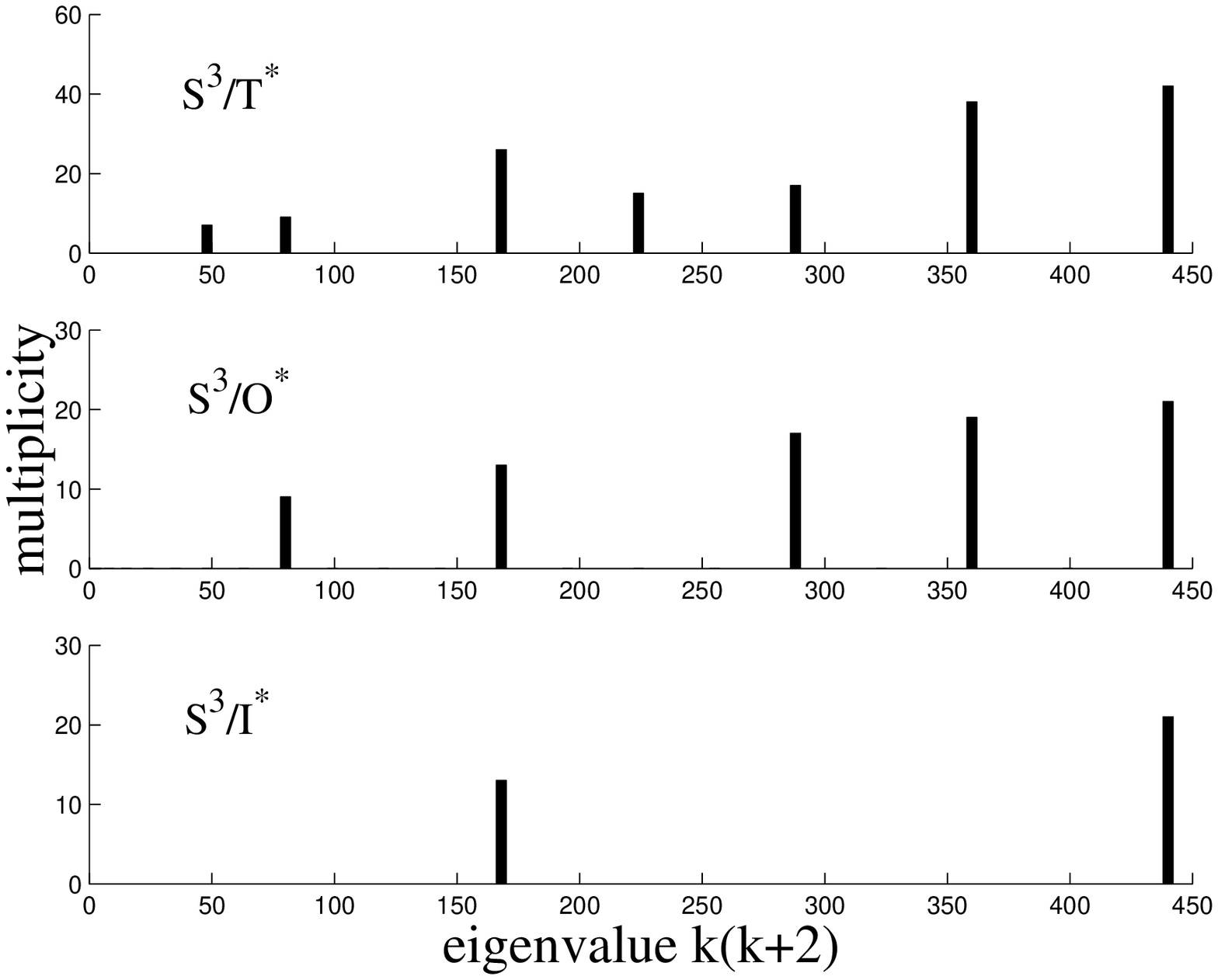,width=7cm} \caption{Spectra
of single-action manifolds generated from the binary tetrahedral,
octahedral and icosahedral groups.} \label{fig3}
\end{center}
\end{figure}

More related to our purpose is the work by Ikeda~\cite{Ikeda95}
in which the spectra of single-action manifolds are determined.  This
result is of great interest for comparison with our numerical
computation.  Unfortunately, it will give us only the wavenumbers with
their multiplicity for a restricted set of topologies and it does not
determine the eigenfunctions.  In tables~\ref{table1} and
\ref{table2}, we sum up the main results on the wavenumbers of
single-action manifolds.  In figure~\ref{fig1}, we compare the spectra
of the projective space $P^3 = S^3/Z_2$ and of the 3-sphere.  Due to
the
$Z_2$ symmetry, half of the modes are lost because they are not
invariant under the antipodal map.  In figure~\ref{fig2} we compare
the spectra of
the single-action manifolds obtained from the groups $Z_7$, $Z_8$,
$Z_9$, $D^*_2$, $D^*_3$, and $D^*_4$ and figure~\ref{fig3} compares
the ones obtained from the binary tetrahedral, octahedral and
icosahedral groups.

\begin{table}
\begin{center}
\begin{tabular}{|l|ll|}
\hline
Manifold             &  Eigenvalue $k(k+2)$  & multiplicity \\
\hline
$S^3/Z_{2p}$   & $k>0$ even & $(k+1)\sum_{l=0}^{k}n_{kl}$ \\
$S^3/Z_{2p+1}$ & $k\geq 2p+1$ odd & $(k+1)\sum_{l=0}^{k}n_{kl}$ \\
               & $k>0$ even & \\
$S^3/D^*_m$    & $\kb$ even & $(2\kb+1)([\kb/m]+1)$ \\
               & $\kb>m$ odd & $(2\kb+1)[\kb/m]$ \\
$S^3/T^*$      & $\kb\not=1,2,5$  & $(2\kb+1)(1+2[\kb/3]+[\kb/2]-\kb)$\\
$S^3/O^*$      & $\kb\not=1,2,3,5,7,11$&
$(2\kb+1)(1+[\kb/4]+[\kb/3]+[\kb/2]-\kb)$\\
$S^3/I^*$      & $\kb\not=1,2,3,4,5,7,8,$&
$(2\kb+1)(1+[\kb/5]+[\kb/3]+[\kb/2]-\kb)$ \\
               & $\quad\quad 9,11,13,14,17,19,23,29$ & \\
\hline
\end{tabular}
\end{center}
\caption{Spectra of single-action manifolds.  For spaces other than
$S^3/Z_n$ we have set $k=2\kb$ with $\kb$ being an integer.  $[p]$
refers to the integer part of $p$ and $n_{kl}$ is defined by
$n_{kl}=1$ if $k\equiv2l({\rm mod}\,n)$ and 0 otherwise.  Adapted
from
Ref.~\cite{Ikeda95}.}
\label{table1}
\end{table}

\begin{table}
\begin{center}
\begin{tabular}{|l|c|c|c|}
\hline
Manifold & First eigenvalue $k(k+2)$ & $k_{_{\rm min}}$ & multiplicity \\
\hline
$S^3/Z_2$              &    8    &   2 &    9  \\
$S^3/Z_n$, $n>2$       &    8    &   2 &    3  \\
$S^3/D_{2}^*$          &   24    &   4 &   10  \\
$S^3/D_{m}^*$ , $m>2$  &   24    &   4 &    5  \\
$S^3/T^*$              &   48    &   6 &    7  \\
$S^3/O^*$              &   80    &   8 &    9  \\
$S^3/I^*$              &  168    &  12 &   13  \\
\hline
\end{tabular}
\end{center}
\caption{Value and multiplicity of the first non zero eigenvalue for
the
single-action manifolds. From Ref.~\cite{Ikeda95}. Note that we have
corrected a mistake on the multiplicity of the first eigenvalue of
$S^3/D_{m}^*$.}
\label{table2}
\end{table}

\section{Numerical determination of the eigenmodes}
\label{sec_num}

To determine the eigenvalues and eigenmodes of the Laplacian, one has
to find a way to take into account the boundary conditions imposed by
the topology.  Different routes have been investigated.  The case of
locally Euclidean manifolds is somehow trivial since the problem can
be solved analytically (see e.g.~\cite{levin98b}).  The hyperbolic
case was first addressed using the boundary element method first
developed by Aurich and Steiner~\cite{aurich89} for the study of
2-dimensional hyperbolic surfaces.  Inoue~\cite{inoue99} developed the
direct boundary element method and was the first to determine precise
eigenmodes of 3-dimensional compact hyperbolic manifolds and get the
36 first eigenmodes of Thurston space for $k\leq 10$~\cite{inoue99}
and then for $k\leq 13$~\cite{inoue00b} (see also Ref.
~\cite{aurich96} for the first computation of the eigenmodes of a cusp
manifold).  Recently a new method was proposed by Cornish and Spergel
\cite{cornish99} in the framework of hyperbolic spaces; such a method
can be adapted to the case of spherical spaces, and thus will be
described in more details below as the ``ghosts method".

In this section, we present three independent methods to compute the
eigenmodes.

\subsection{Ghosts method}

The ghosts method is based on the idea that any square integrable
function
in $L^2(X/\Gamma)$, $X$ being the universal covering space, satisfies
\begin{equation}\label{c1}
\Psi(x)=\Psi(gx)
\end{equation}
for all $g\in\Gamma$. Any function of $L^2(X/\Gamma)$ can be
lifted to a ($\Gamma$-invariant) function of $L^2(X)$ and,
reciprocally, any $\Gamma$-invariant function projects down to a
function on $X/\Gamma$. It follows that the eigenmodes of the
Laplacian can be decomposed as
\begin{equation}\label{c3}
\Psi_{\beta,s}^{^{[\Gamma]}}(x)=\sum_{\ell=0}^\infty\sum_{m=-\ell}^{\ell}
\xi_{\beta,s\ell m}{\cal Y}_{\beta\ell m}(x)
\end{equation}0
where there is no summation on $\beta$ because the eigenfunctions of
the universal covering space, ${\cal Y}_{\beta\ell
m}=R_{\beta\ell}(\chi)Y_{\ell m}(\theta,\phi)$ (see~\ref{appA}) form a
complete and linearly independent family.  Note that the coefficients
$\xi_{\beta,s\ell m}$ are obtained once the holonomies are known and
they will depend on the base point.

Now, choose randomly $d$ points, $x_i$, in the fundamental domain
and consider the $n_i$ images of each point up to a distance
$\rho_{_{\rm max}}$. We also need to truncate the sum (\ref{c3})
to a maximum value $L$ of $\ell$. Each point generates
$n_i(n_i+1)/2$ constraints of the form (\ref{c1})
\begin{equation}\label{c2}
\Psi_{\beta,s}^{^{[\Gamma]}}(g_a x_i)-\Psi_{\beta,s}^{^{[\Gamma]}}(g_b
x_i)=0
\end{equation}
for $a\not=b$ and $a,b=\ldots\vert\Gamma\vert$. With the
decomposition (\ref{c3}) the set of constraints (\ref{c2}) for all
the $d$
points $x_i$ takes the form
\begin{equation}\label{systeme}
{\bf A}_\beta v_{\beta,s}=0
\end{equation}
where the $(L+1)^2$-components vector $v_{\beta,s}$ is defined by
\begin{equation}
v_{\beta a}\equiv\left[ \begin{array}{c} \xi_{\beta,s00} \\ \vdots \\
\xi_{\beta,sLL}
\end{array}\right]
\end{equation}
The matrix ${\bf A}_\beta$ with $M=\sum_{j=1}^dn_j(n_j+1)/2$ rows and
$N=(L+1)^2$ columns is defined by\footnote{Note that there is a
typo in the equation (2.3) of \cite{cornish99}.}
\begin{equation}
{\bf A}_\beta\equiv\left[
\begin{array}{ccc}
{\cal Y}_{\beta00}(g_{1}x_{1})-{\cal Y}_{\beta00}(g_{2}x_{1}) & \ldots
& {\cal Y}_{\beta LL}(g_{1}x_{1})-{\cal Y}_{\beta LL}(g_{2}x_{1}) \\
\vdots & \ddots & \vdots\\ {\cal Y}_{\beta00}(g_ax_{d})-{\cal
Y}_{\beta00}(g_bx_{d}) & \ldots & {\cal Y}_{\beta LL}(g_ax_{d})-{\cal
Y}_{\beta LL}(g_bx_{d})
\end{array}
\right]
\end{equation}
If $M>N$ the system (\ref{systeme}) is over constrained and has a
solution if and only if $q$ is an eigenvalue of the compact space.  We
named this method as the ghosts method because of its close analogy
with the simulation of galaxy catalogs in a multi-connected
universe~\cite{us3,us0,us1,us2}.

Numerically, one uses a single value decomposition (SVD) method to
extract the eigenmodes, i.e. which form a basis of Ker(${\bf
A}_\beta$). This decomposition is based on the theorem stating that
any $M\times N$ matrix ${\bf A}$ with $M\geq N$ can be written as the
product
\begin{equation}
{\bf A}={\bf UD} {}^t{\bf V}
\end{equation}
where ${\bf U}$ is a $M\times N$ unitary matrix, ${\bf D}={\rm
diag}(w_1,\ldots,w_N)$ is a $N\times N$ diagonal matrix, the $w_i$
being
non negative and ${\bf V}$ is a $N\times N$ unitary matrix.  The
columns
of ${\bf U}$ associated to non-zero $w_i$ form an orthonormal
basis spanning the range of ${\bf A}_\beta$. The columns of ${\bf V}$
corresponding to the vanishing $w_i$ form a basis of the nullspace of
${\bf A}_\beta$ since it is solution of the equation (\ref{systeme}),
i.e. of the subspace $E_\beta^{^{[\Gamma]}}$ of the eigenmodes of
$X/\Gamma$ with eigenvalue $q$.

This method was first implemented~\cite{cornish99} to compute the
lowest eigenvalues and eigenfunctions for 12 hyperbolic manifolds.
We have extended the method to include any topology. In hyperbolic
spaces, we recover the results of \cite{cornish99} and in the
Euclidean case we compared our result with the analytic solutions.
When the universal covering space is not compact, one has to
specify the value of the degree of constraint $c=M/N$, of the
cut-off $L$ and of the maximum radius $\rho_{_{\rm max}}$ up to
which the images are considered. Our main interest in this article
is the case of spherical spaces and it enjoys a number of
simplifications. First, we can set $\rho_{_{\rm max}}=\pi$ since
the group $\Gamma$ is finite. It follows that any of the $d$
points has exactly $\vert\Gamma\vert-1$ images so that
$M=d\vert\Gamma\vert(\vert\Gamma\vert-1)/2$ and we end with only
two free parameters $(L,d)$ or equivalently $(L,c)$. This
simplifies the discussion concerning the choice of the different
cut-offs $L$.

The parameter $c$ is adjusted numerically. It needs to increase fast
with the order $|\Gamma|$ of the group, but for each order $c$ needs
to increase slowly with the increase of $\beta$. For example for
$|\Gamma|=8$, $c=|\Gamma|+\beta$ works very well until $\beta=15$. For
$\beta>15$, we need either a smaller $c$ or a $c$ that increases
slower with $\beta$.  The constraint determines the $d$ points that we
need to generate randomly in the fundamental domain for each
$\beta$. Apparently the only reason for choosing a specific value for
$d$ is that the method does not work without a good balance between
$M$ (number of rows) and $N$ (number of columns) for each $|\Gamma|$
and $\beta$.  A wrong choice of $c$ could cause the failure of the
method. The results of this method have been compared with Ikeda's
results and agree with them.

\subsection{Averaging method}

When focusing on spherical spaces, one can take into account the fact
that the number of images of any point is exactly given by the order
of the group, and is thus finite, to develop another numerical method
for computing the eigenfunctions, completely independent from the
previous one.

This method is based on the two remarks that
\begin{itemize}
\item if $\Psi$ is an eigenmode, then $\Psi\circ g$ is also an
eigenmode
\item  for any function $f$ on $S^3$, then $\bar f$ defined
by
\begin{equation}\label{averaging_formula}
\bar f=\frac{1}{\vert\Gamma\vert}\sum_{g\in\Gamma}f\circ g
\end{equation}
is a $\Gamma$-invariant function of $L^2_\Gamma(S^3)$, the indice
recalling the $\Gamma$-invariance, which can thus be identified to a
function of $L^2(S^3/\Gamma)$. The operation
\begin{equation}\label{defav}
{\rm av}_\Gamma:\left\lbrace
\begin{array}{l}
L^2(S^3)\rightarrow L^2_\Gamma(S^3)\\
f \mapsto\bar f
\end{array}\right.
\end{equation}
is a projection on the subspace of $\Gamma$-invariant functions (since
${\rm av}_\Gamma^2=1$ and ${\rm av}_\Gamma(f)=f$ iff $f$ is
$\Gamma$-invariant). This induces an equivalence relation $g\sim f$
iff ${\rm av}_\Gamma(g)={\rm av}_\Gamma(f)$ so that $L^2(S^3/\Gamma)$
is just given as the set of the functions of $L^2(S^3)$ modulo the
equivalence relation, i.e. $\{{\rm av}_\Gamma(f),\,f\in L^2(S^3)\}$
\end{itemize}
It follows that the eigenmodes of the Laplacian on
$S^3/\Gamma$ are explicitly given in terms of the eigenmodes
of the Laplacian on the universal covering space (see~\ref{appA})
by
\begin{equation}\label{famille}
\Psi^{^{[\Gamma]}}_{\beta,s}=
\frac{1}{\vert\Gamma\vert}\sum_{g\in\Gamma}{\cal Y}_{\beta\ell
m}\circ g.
\end{equation}
Indeed, the $(k+1)^2$ linearly independent eigenmodes ${\cal
Y}_{k\ell m}$ project down to $(k+1)^2$ $\Gamma$-invariant
eigenmodes $\Psi^{^{[\Gamma]}}_{k}$.  This set of functions is a
generator of the eigenspace $E_k^{^{[\Gamma]}}$ but it is not a free
family because we expect ${\rm dim}(E_k^{^{[\Gamma]}})<(k+1)^2$.  We
will thus have to pick up the ${\rm dim}(E_k^{^{[\Gamma]}})$
independent functions of the family (\ref{famille}).  This can be
performed by using an orthonormalisation procedure (the classical
Gram-Schmidt method itself being numerically disastrous).

Numerically, we compute the average family (\ref{famille}) and
then decompose it on the basis ${\cal Y}_{k\ell m}$ as
\begin{equation}
\Psi^{^{[\Gamma]}}_{\beta,s}=\sum_{\ell,m}\xi_{\beta,s\ell m}{\cal
Y}_{\beta\ell m}
\end{equation}
and use the SVD method (as described in the previous section) to
perform the orthonormalisation. For that purpose we have to choose
$d=(k+1)^2$ and $L=\beta$ so that there is no free parameter to
choose and
\begin{equation}
c=\frac{1}{2}\left(\frac{k+1}{k+2}\right)^2|\Gamma|(|\Gamma|-1).
\end{equation}

\subsection{Projection method}\label{subsec_projection}

The ghosts method and the averaging method, presented in the two
previous sections, both compute an orthonormal basis
$\Psi^{^{[\Gamma]}}_{\beta\ell m}$ for the space of eigenmodes of
$S^3/\Gamma$, so that we may later compute a random eigenmode
$\Psi^{^{[\Gamma]}}_{\beta}$ of $S^3/\Gamma$ as a linear combination
(\ref{c3}).  The projection method, by contrast, computes the random
eigenmode $\Psi^{^{[\Gamma]}}_{\beta}$ directly, without explicitly
computing a basis for the eigenspace.  Throughout this section we
assume the wavenumber $k$ is fixed, thus fixing the eigenvalue
$k(k+2)$ and the parameter $\beta = k + 1$ as well.

\subsubsection{The algorithm}\label{ProjectionAlgorithm}

The idea is to first compute a random eigenmode of $S^3$,
and then project that random eigenmode down to an eigenmode
of $S^3/\Gamma$.  More precisely, one proceeds as follows:
\begin{description}
\item[Step 1:]  Construct a random eigenmode on $S^3$.

    Construct a random eigenmode $\Psi_{\beta}$ on $S^3$
    as a linear combination
      \begin{equation}\label{jeff1}
        \Psi_{\beta}=\sum_{\ell,m}\zeta_{\beta \ell m}
        {\cal Y}_{\beta\ell m}.
       \end{equation}
    Choose the coefficients $\zeta_{\beta\ell m} $ relative
    to a Gaussian distribution with mean 0 and standard deviation 1.
    The resulting point
    $\left(\zeta_{\beta\ell m} \right) \in {\bf R}^{\beta^2}$ will,
    in effect, be chosen relative to a spherically symmetric
    distribution, because the product measure
    $$
      \prod_{\ell,m}\exp(-\zeta_{\beta\ell m}^2/2)=
      \exp(-r^2/2)
      $$
    depends only on the radial distance
    $r^2 = \sum_{\ell,m}\zeta_{\beta\ell m}^2 $ in ${\bf
R}^{\beta^2}$.
    Note that the expected value $E(\zeta_{\beta\ell m})$ of each
coefficient
    $\zeta_{\beta\ell m}$ is 1, so the expected value of the squared
radius $r^2$
    is simply the dimension $\beta^2$ of the space:
    $$E(r^2) = E(\sum_{\ell,m}\zeta_{\beta\ell m}^2) =
    \sum_{\ell,m}E(\zeta_{\beta\ell m}^2) = \sum_{\ell,m}1 =
\beta^2.$$
    In the infinitely unlikely event that $r = 0$, discard
    the randomly chosen $\zeta_{\beta\ell m} $ and choose a new set.

\item[Step 2.]  Construct the average ${\rm av}_\Gamma(\Psi_{\beta})$.

    Given the eigenmode $\Psi_{\beta}$ of $S^3$, define the
$\Gamma$-invariant
    eigenmode ${\rm av}_\Gamma(\Psi_{\beta})$ of $S^3/\Gamma$
    via the averaging formula (\ref{averaging_formula}).
    As mentioned earlier, a $\Gamma$-invariant eigenmode of $S^3$
    corresponds to an eigenmode of the quotient space $S^3/\Gamma$.
    In principle we can evaluate formula (\ref{averaging_formula})
    for any $x \in S^3$, but in practice we need evaluate it only
    for those $x$ lying on the last scattering surface.

    {\it Computational note:}  We begin with $x$ in rectangular
coordinates
    $(x_1,x_2,x_3,x_4)$, so that $g(x)$ may be computed quickly and
    easily as a $4\times 4$ matrix times a 4-element vector.
    We then convert the result to spherical or toroidal coordinates
    $(\chi,\theta,\varphi)$ for efficient evaluation of
Eq.~(\ref{jeff1}).
\end{description}

\subsubsection{Proof of orthogonality}

Let $n = \beta^2$ be the dimension of the full $\beta$-eigenspace
of $S^3$, and let $m$ be the (unknown) dimension of the
$\Gamma$-invariant
subspace, i.e. $m$ is the dimension of the $\beta$-eigenspace of
$S^3/\Gamma$.  Thus the averaging operator ${\rm av}_\Gamma()$ in Step 2
projects the eigenspace ${\bf R}^n$ of $S^3$ down onto the subspace
${\bf R}^m$ of $\Gamma$-invariant eigenmodes.  The question is,
relative to the standard inner product in function space,
is this an orthogonal projection from ${\bf R}^n$ to ${\bf R}^m$?
If so, then a spherically symmetric distribution of points in
${\bf R}^n$ (corresponding to random eigenmodes of $S^3$) will
project to a spherically symmetric distribution of points in
${\bf R}^m$ (corresponding to random eigenmodes of $S^3/\Gamma$).
However, if the projection is not orthogonal -- for example if
it involves a shearing motion -- then a spherically symmetric
distribution of points in ${\bf R}^n$ will {\it not} project
to a spherically symmetric distribution in ${\bf R}^m$, but
rather to a distribution that is slanted to one side or another.
Fortunately the projection is orthogonal, as shown in Proposition I.

\vskip0.25cm

\noindent Proposition I.  {\it Let $H$ be the space of all
$\beta$-eigenmodes on $S^3$, and $H_\Gamma$ be the space of
$\beta$-eigenmodes that are invariant under the action of a finite
group $\Gamma \subset SO(4)$.  Then the averaging function ${\rm
av}_\Gamma$ defined in Eq.~(\ref{defav}) maps $H$ orthogonally onto
$H_\Gamma$.}

\vskip0.25cm

\noindent {\it Proof.}
Let $H_\perp$ be the orthogonal complement of $H_\Gamma$ in $H$.
That is, let $H_\perp$ consist of those eigenmodes in $H$ that
are orthogonal to all eigenmodes in $H_\Gamma$.
Thus $H \simeq {\bf R}^n$, $H_\Gamma \simeq {\bf R}^m$, and
$H_{\perp} \simeq {\bf R}^{n-m}$ for some $n$ and $m$.
Our goal is to show that the averaging function ${\rm av}_\Gamma$
maps $H_\perp$ to 0.

Let $\Psi_\perp$ be an element of $H_\perp$.  This means that
$\left<\Psi_\perp, \Psi_\Gamma\right> = 0$ for all $\Psi_\Gamma \in
H_\Gamma$.
The inner product $\left<,\right>$ is invariant under the action
of $SO(4)$, so for every $g \in \Gamma$,
$\left<\Psi_\perp \circ g, \Psi_\Gamma \circ g\right> = 0$.
But $\Psi_\Gamma$ is, by definition, invariant under $\Gamma$, so
$\Psi_\Gamma \circ g = \Psi_\Gamma$, hence
$\left<\Psi_\perp \circ g, \Psi_\Gamma\right> = 0$ for
all $\Psi_\Gamma \in H_\Gamma$.  In other words,
$\Psi_\perp \circ g \in H_\perp$ for all $g \in \Gamma$.
This implies that ${\rm av}_\Gamma(\Psi_\perp) \in H_\perp$.
But ${\rm av}_\Gamma(\Psi_\perp)$ is also an element of $H_\Gamma$.
Because $H_\Gamma \cap H_\perp = 0$, this implies
${\rm av}_\Gamma(\Psi_\perp) = 0$, as required.  {\it QED}

\subsubsection{Interpreting the norm}  The squared norm
$|{\rm av}_\Gamma(\Psi_\beta)|^2$ provides a statistical estimate
of the dimension of the $\Gamma$-invariant eigenspace.

\vskip0.25cm

\noindent Proposition II.  {\it For each $\beta$, the expected value
of the squared norm $\left< {\rm av}_\Gamma(\Psi_\beta),
{\rm av}_\Gamma(\Psi_\beta) \right>$
is the dimension of the $\Gamma$-invariant eigenspace $H_\Gamma$.}

\vskip0.25cm

\noindent {\it Proof.}
The comments in Step 1 of the algorithm in Section
\ref{ProjectionAlgorithm}
show that the expected squared norm of the original $\Psi_\beta$
(before averaging) tells the dimension of the full eigenspace:
$$E(\left<\Psi_\beta, \Psi_\beta\right>) = E(\sum{\zeta_{\beta\ell
m}^2}) = \beta^2.$$
Choose an orthonormal basis $\{{\cal Y}_i\}$ for the eigenspace
such that the basis vectors $\{{\cal Y}_1,\dots,{\cal Y}_m\}$ span the
$\Gamma$-invariant eigenspace $H_\Gamma$ while the remaining basis
vectors
$\{{\cal Y}_{m+1},\dots,{\cal Y}_n\}$ lie orthogonal
to $H_\Gamma$.
Because the basis $\{{\cal Y}_i\}$ is orthonormal,
the distribution $\exp(-r^2 / 2)$ factors as a product
$$\exp(-r^2 / 2) =  \prod_{i=1}^n \exp(-\zeta_i^2 / 2).$$
Proposition I implies that, relative to this basis, the averaging
operator ${\rm av}_\Gamma()$ preserves the first $m$ coordinates while
collapsing the remaining $n - m$ coordinates to zero.
Thus the distribution of ${\rm av}_\Gamma(\Psi_\beta)$ is given
by the restricted product
$$\prod_{i=1}^m \exp(-\zeta_i^2 / 2) = \exp(-r_\Gamma^2 / 2)$$
where $r_\Gamma^2 = \sum_{i=1}^m{\zeta_i^2}$, and the above reasoning
now implies that
$$E(\left<{\rm av}_\Gamma(\Psi_\beta), {\rm av}_\Gamma(\Psi_\beta)\right>)
= m,$$
as required.
{\it QED}

\vskip0.25cm

Proposition II remains valid even if we do not explicitly construct
the basis $\{{\cal Y}_i\}$.  We may instead compute the squared norm
$\left<{\rm av}_\Gamma(\Psi_\beta), {\rm av}_\Gamma(\Psi_\beta)\right>$ by
sampling points.
Taking 32 random eigenmodes $\Psi_\beta$ and for each one evaluating
$\left<{\rm av}_\Gamma(\Psi_\beta), {\rm av}_\Gamma(\Psi_\beta)\right>$ at
32 random points yields the following rough estimates for the
dimension of the eigenspace (PDS designates the Poincar\'e
Dodecahedral Space of order 120):
\begin{center}
\begin{tabular}{|l|rrrrrrrrrrr|}
\hline
$k$  & 2 & 3 & 4 & 5 & 6 & 7 & 8 & 9 & 10 & 11 & 12 \\
\hline
$S^3$  & 9.1 & 17.1 & 22.2 & 40.0 & 44.1 & 60.7 & 83.9 & 106.4 & 129.5 &
144.6 & 163.9 \\
$P^3$  & 8.5 &  0.0 & 23.9 &  0.0 & 47.2 &  0.0 & 80.5 &   0.0 & 120.2 &
0.0 & 165.8 \\
L(5,2) & 1.4 &  3.6 &  3.9 &  6.6 &  7.9 & 12.7 & 16.5 &  19.0 &  26.9 &
25.1 &  31.9 \\
PDS    & 0.0 &  0.0 &  0.0 &  0.0 &  0.0 &  0.0 &  0.0 &   0.0 &   0.0 &
0.0 &  11.8 \\
\hline
\end{tabular}
\end{center}
The above computations took only a few minutes on a 300 MHz desktop
computer.
Increasing the number of random eigenmodes and the number of sampling
points would increase the accuracy of the results at the expense of
a longer computation time.

\subsubsection{Computational complexity}

The projection method is reasonably fast.
The choice of the $\zeta_{\beta l m}$ in Step 1 requires
only $O(\beta^2)$ time, and is completed almost
instantaneously on a desktop PC.
Thereafter the evaluation of ${\rm av}_\Gamma(\Psi_{\beta})(x)$
for each point $x \in S^3$ takes $\vert\Gamma\vert$
times as long as evaluating the underlying random eigenmode
$\Psi_{\beta}(x)$ of $S^3$ at the same point $x$.
In other words, using the projection method,
an eigenmode of $S^3/\Gamma$ is $\vert\Gamma\vert$ times
as expensive to compute as an eigenmode of $S^3$.
More precisely, the time to evaluate $\Psi_{\beta}(x)$
grows as $|\Gamma|\,\beta^3,$
because for a fixed value of $\beta$ the eigenspace has
dimension $\beta^2$, meaning that there are $\beta^2$
terms to evaluate, each of which requires $O(\beta)$ steps.
In practice we evaluated the eigenmodes of $S^3$ using
the toroidal coordinates method of \cite{wlu02}, but in
principle the same runtime could be obtained using spherical
coordinates.

\section{Analytical solutions for lens and prism spaces}
\label{sec_torus}

Besides the numerical methods presented above, there are special cases
for which the eigenfunctions can obtained analytically~\cite{wlu02}.
The results are indeed of importance to test the accuracy of our
numerical computations.  The method is based on the use of torus
coordinates and applies to lens $L(p,q)$ and prism $S^3/D_m^\ast$
spaces.  We briefly recall the main points and results
of~\cite{wlu02}.

\subsection{Summary of the general method}

The key of the method is to choose a coordinate system that
respects the holonomy group $\Gamma$. We introduce the coordinates
in ${\bf R}^4$, $(x, y, z, w)$ by
\begin{eqnarray}\label{CoordinateDefinition}
x &=& \cos{\chi'}\,\cos{\theta'} \nonumber\\ y
&=&\cos{\chi'}\,\sin{\theta'} \nonumber\\ z &=&
\sin{\chi'}\,\cos{\varphi'}\nonumber\\ w &=&
\sin{\chi'}\,\sin{\varphi'}
\end{eqnarray}
so that the equation of the 3-sphere is simply
$x^2+y^2+z^2+w^2=1$. Note that they are different from the
4-dimensional coordinates (\ref{klein}) introduced previously and
that now the intrinsic coordinates have to range as
\begin{eqnarray}
0\leq\chi'\leq \pi/2, \qquad 0 \leq\theta'\leq2\pi \qquad 0\leq
\varphi'\leq 2\pi.
\end{eqnarray}
For each fixed value of $\chi' \in [0, \pi/2]$, the $\theta'$ and
$\varphi'$ coordinates sweep out a torus.  Taken together, these
tori almost fill $S^3$.  The exceptions occur at the endpoints
$\chi' = 0$ and $\chi' = \pi/2$, where the stack of tori collapses
to the circles $x^2 + y^2 = 1$ and $z^2 + w^2 = 1$, respectively.

Identifying the eigenmodes of $S^3/\Gamma$ and the
$\Gamma$-invariant eigenmodes of $S^3$, as explained in our
introductory remark, the eigenmodes of a lens or prism space are
given by $Z_p$-invariant or $D_m^*$-invariant eigenmodes of $S^3$.

An elementary construction~\cite{wlu02} shows that for each
wavenumber $k$, with eigenvalue $k(k+2)$, the corresponding eigenspace
of $S^3$ is spanned by the basis
\begin{equation}\label{Eigenbasis}
B_k = \{\,{\cal Q}_{k\ell m}\,|\quad|\ell| + |m| \leq k\quad{\rm
and} \quad|\ell| + |m| \equiv k\;{\rm (mod\,2)}\,\}
\end{equation}
where
\begin{eqnarray} \label{PsiSolution}
{\cal Q}_{k\ell m} &=&
\cos^{|\ell|}\chi'\,\sin^{|m|}\chi'\,P^{|m|,|\ell|}_{d}(\cos 2\chi')
\nonumber\\
&\times& (\cos{|\ell| \theta' }\quad{\rm or}\quad \sin{|\ell| \theta' })
\times (\cos{|m| \varphi'}\quad{\rm or}\quad \sin{|m| \varphi'})
\end{eqnarray}
$P^{|m|,|\ell|}_{d}$ being the Jacobi polynomial
$$P^{|m|,|\ell|}_{d}(x) = \frac{1}{2^d} \sum_{i=0}^d {{|m|+d}\choose{i}}
{{|\ell|+d}\choose{d-i}}\,(x + 1)^i\,(x - 1)^{d-i}$$
and $\cos{|\ell| \theta' }$ (resp.  $\sin{|\ell| \theta' }$) being used
when $\ell \geq 0$ (resp.  $\ell < 0$), and similarly for the choice
of $\cos{|m| \varphi'}$ or $\sin{|m| \varphi'}$.

It is straightforward to see how the generating isometry of a lens
space $L(p,q)$, given in rectangular coordinates by
\begin{equation}
\left(
\begin{array}{cccc}
  \cos 2\pi/p & -\sin 2\pi/p &     0       &      0       \\
  \sin 2\pi/p &  \cos 2\pi/p &     0       &      0       \\
      0      &      0      & \cos 2\pi q/p & -\sin 2\pi q/p \\
      0      &      0      & \sin 2\pi q/p &  \cos 2\pi q/p \\
\end{array}
\right) \label{IsometryRectangular}
\end{equation}
or in toroidal coordinates by
\begin{eqnarray}\label{IsometryTorus}
\chi'    &\to& \chi' \nonumber \\ \theta'  &\to& \theta'  + 2\pi/p
\nonumber \\ \varphi' &\to& \varphi' + 2\pi q/p,
\end{eqnarray}
acts on the ${\cal Q}_{k\ell m}$.  The eigenmodes of $L(p,q)$
comprise the fixed point set of this action. A set of simple
numerical conditions tells how to select an orthogonal basis for
this fixed point set, essentially as a subset of the basis $B_k$.
Specifically, the eigenbasis for $L(p,q)$ includes
\begin{equation}\label{LensSpaceConditions}
\Psi_{k\ell m}^{{[L(p,q)]}}=\left\lbrace
\begin{array}{ll}
{\cal Q}_{k00} & {\rm always} \\
{\cal Q}_{k\pm\ell0} & {\rm iff}\, \ell \equiv 0 (\mathrm{mod} p) \\
{\cal Q}_{k0\pm m}& {\rm iff}\, qm \equiv 0 (\mathrm{mod} p) \\
\frac{{\cal Q}_{k\ell m}+{\cal Q}_{k-\ell-m}}{\sqrt{2}},
\frac{{\cal Q}_{k-\ell m}-{\cal Q}_{k\ell-m}}{\sqrt{2}} & {\rm
iff}\, \ell \equiv qm (\mathrm{mod} p) \\
\frac{{\cal Q}_{k\ell m}-{\cal Q}_{k-\ell-m}}{\sqrt{2}},
\frac{{\cal Q}_{k-\ell m}+{\cal Q}_{k\ell-m}}{\sqrt{2}}
   & {\rm iff}\, \ell \equiv -qm (\mathrm{mod} p)
\end{array}
\right.
\end{equation}
For details as well as for the explicit form of the eigenmodes of
prism spaces, please see Ref.~\cite{wlu02}.  A similar analysis yields
an explicit eigenbasis for a prism space.  The ${\cal Q}_{k\ell m}$
are already mutually orthogonal, so after normalizing them to unit
length we may use the above basis to construct unbiased random
eigenmodes of a lens or prism space, with wavenumber $k$.

\subsection{Extracting the coefficients $\xi_{\beta a\ell m}$}

The previous analysis gives the decomposition of the eigenmodes on the
basis ${\cal Q}_{\beta\ell m}$ as
\begin{equation}\label{2}
\Psi_{\beta,s}=\sum\eta_{\beta,s\ell m}{\cal Q}_{\beta\ell
m}(\chi',\theta',\varphi')
\end{equation}
but what we need are the coefficients $\xi_{\beta,s\ell m}$ of the
decomposition on the basis ${\cal Y}_{\beta\ell m}$ as given in
Eq.~(\ref{c3}).

${\cal Y}_{\beta\ell m}$ and ${\cal Q}_{\beta\ell m}$ are two basis of
dimension $(k+1)^2$ for each $\beta=k+1$, so up to a change of
coordinates between toroidal and spherical coordinates, we can write
\begin{equation}\label{3}
{\cal Q}_{\beta\ell m}=\sum\alpha_{\beta\ell\ell'mm'}{\cal
Y}_{\beta\ell' m'}.
\end{equation}
Note that $(\ell,m)$ and $(\ell',m')$ do not vary in the same range
since $0\leq\ell'\leq\beta-1$, $|m'|\leq\ell'$ and $|\ell|+|m|<\beta$,
$|\ell|+|m|\equiv\beta-1\,{\rm (mod 2)}$.  The coefficients
$\alpha_{\beta\ell\ell'mm'}$ are explicitly given by
\begin{equation}\label{3bis}
\alpha_{\beta\ell\ell'mm'}=\int{\cal Q}_{\beta\ell m}
(\chi',\theta',\varphi'){\cal Y}^{*}_{\beta\ell' m'}
(\chi,\theta,\varphi)\sin^2\chi\sin\theta \d\chi\d\theta\d\varphi
\end{equation}
where $(\chi',\theta',\varphi')$ are functions of
$(\chi,\theta,\varphi)$. It can be checked from (\ref{klein}) and
(\ref{CoordinateDefinition}) that
\begin{eqnarray}
\chi'&=&{\rm arccos}\left[\cos\chi\cos\theta\right]\\
\theta'&=&{\rm arccos}\left[\frac{\cos\chi\sin\theta}{\sqrt{1-
          \cos^2\chi\cos^2\theta}}\right]\\
\varphi'&=&\varphi.
\end{eqnarray}
It can be checked that the two basis differ by more than just a change
of coordinates.  It follows that for the particular case of lens and
prism spaces, the computation goes as follows:
\begin{enumerate}
\item Determine the coefficients $\eta_{\beta,s\ell m}$ as given in
Ref.~\cite{wlu02}; thus mainly a tedious book-keeping operation.
\item Compute the coefficients of the change of basis (\ref{3}); this
has to be performed once for all for all spaces.
\item The required coefficients are obtained by a matrix
multiplication
\begin{equation}\label{4}
\xi_{\beta,s\ell
m}=\sum_{\ell'm'}\alpha_{\beta\ell\ell'mm'}\eta_{\beta,s\ell'm'}.
\end{equation}
\end{enumerate}

\subsection{The simplest example}

For the projective space, $S^3/Z_2$, there is only one generator which
brings any point to its antipodal point. It follows that the
eigenfunctions are just the average of the spherical harmonics
evaluated with their antipodal equivalent. This can be sorted out
analytically and one easily obtains that
\begin{equation}\label{5}
\xi_{\beta,s\ell m}^{^{[Z_2]}}=\frac{1}{2}\left[1-(-1)^\beta\right]
\delta_{s,(\ell m)}
\end{equation}
where the function $\delta_{s,(\ell m)}=1$ if the label $s$ can be
identified with $(\ell,m)$ and is zero otherwise. We recover the
results from figure~\ref{fig1}, that is that there is no modes for
$\beta$ even and that the dimension of the eigenspace for $\beta$ odd
is equal to the one of $S^3$, that is $(k+1)^2$.

\section{Numerical results}
\label{sec_stat}

An interesting property is the distribution of the number of modes per
wavenumber
interval. In hyperbolic manifolds, the number $N(\leq q)$ of modes
smaller than $q$ is well described by the Weyl asymptotic formula
\begin{equation}\label{weyl}
  N(\leq q)\sim\frac{{\rm Vol}(H^3/\Gamma)}{6\pi^2}(q^2-1)^{3/2}
\end{equation}
for $q\gg1$.  In the case of the 3-sphere, $q^2=k(k+2)$ with
multiplicity mult$(q)=(k+1)^2$ so that
\begin{equation}\label{weyl2}
N^{^{[S^3]}}(\leq
q)=\sum_{p=2}^{k}(p+1)^2=\frac{(k+1)(k+2)(2k+3)-30}{6}
\end{equation}
and the Weyl formula, which applies also to the spherical manifolds,
tells us that
\begin{equation}\label{weyl3}
N^{^{[\Gamma]}}\sim N^{^{[S^3]}}/|\Gamma|.
\end{equation}

We computed the eigenmodes and eigenfunctions of some spherical spaces
with the different methods described above. First, it was checked that
the spectra agree with the theoretical ones in the cases described
in table~\ref{table1}. In the particular case of lens and prism
spaces, the eigenmodes and eigenfunctions agree with the ones obtained
analytically in Ref.~\cite{wlu02}.

The computational time of the averaging method is experimentally found
to be proportional to $|\Gamma|\beta^{4.68}$, the typical running time
being less than 10 seconds on a desktop computer for $\beta<15$.

\begin{figure}
\begin{center}
\epsfig{file=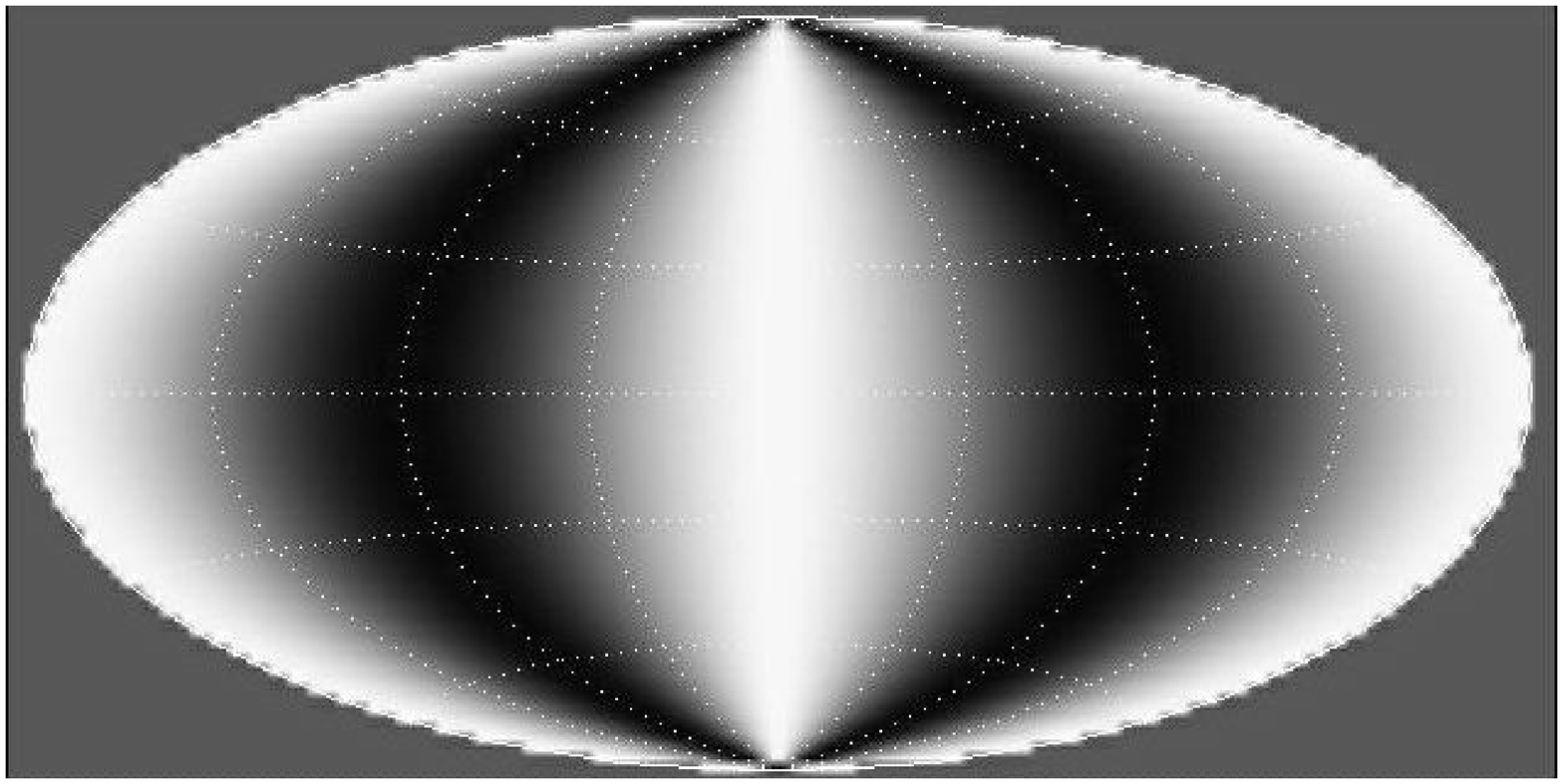,width=7cm}\epsfig{file=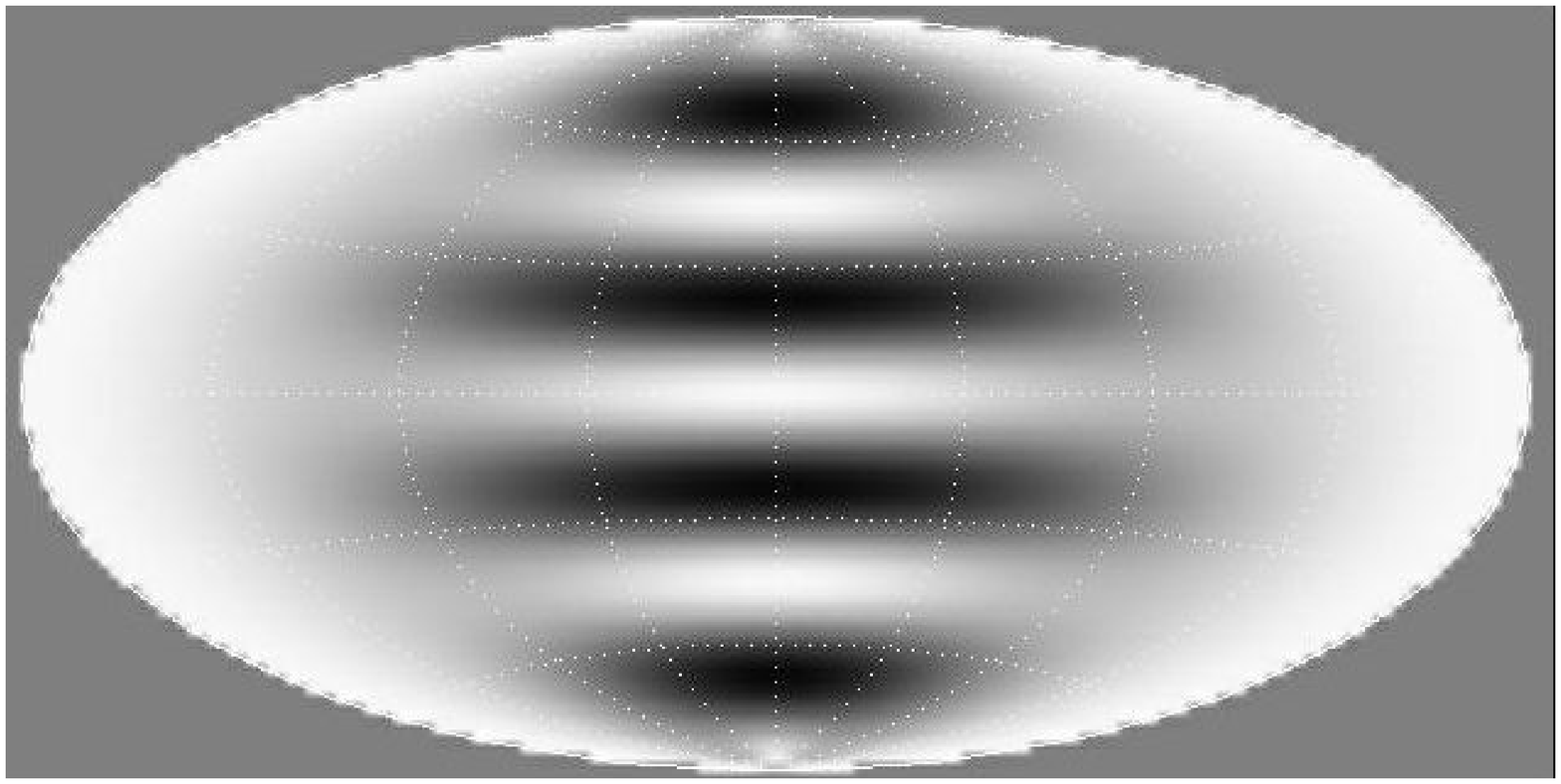,width=7cm}
\epsfig{file=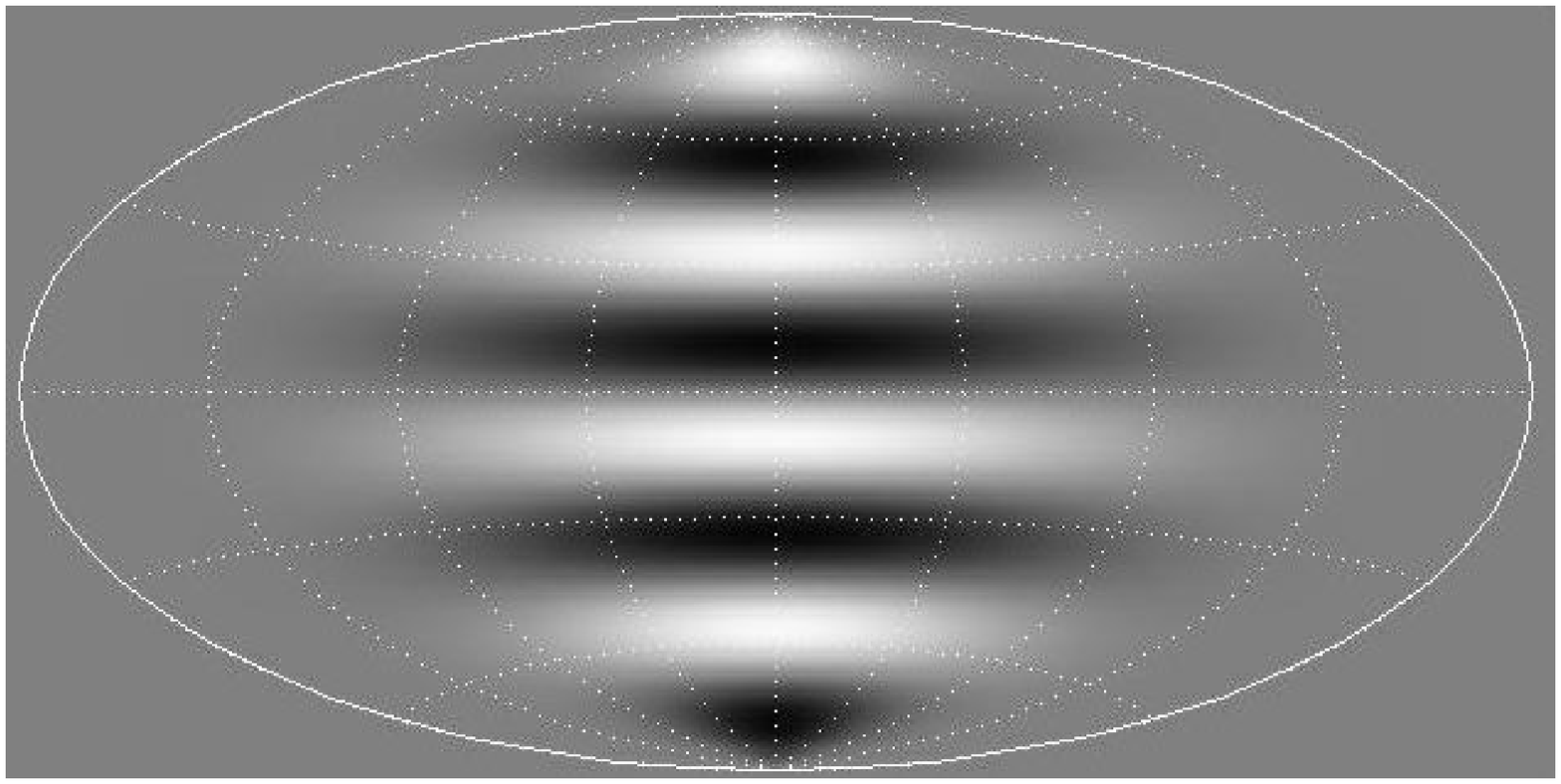,width=7cm}\epsfig{file=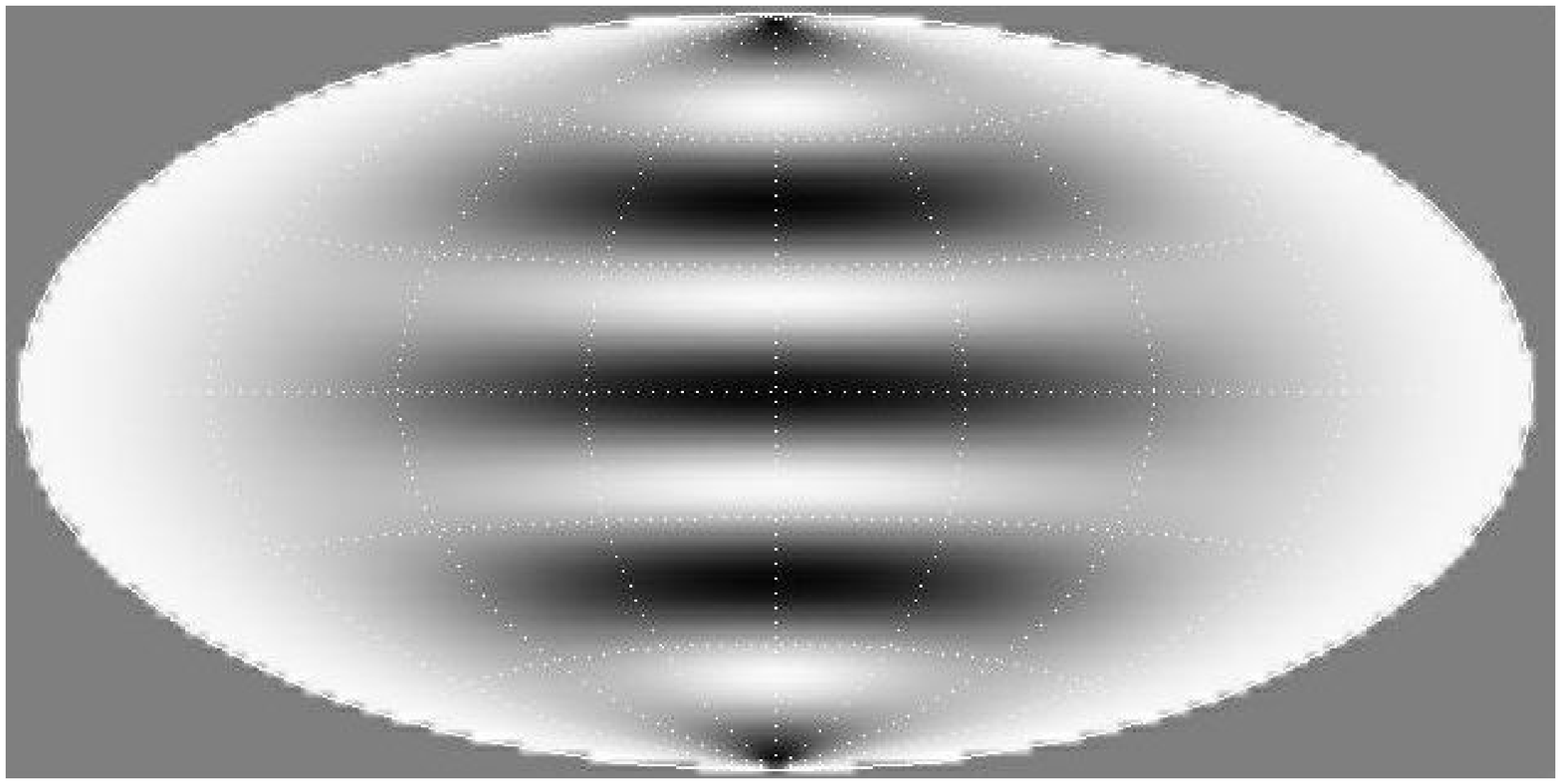,width=7cm}
\epsfig{file=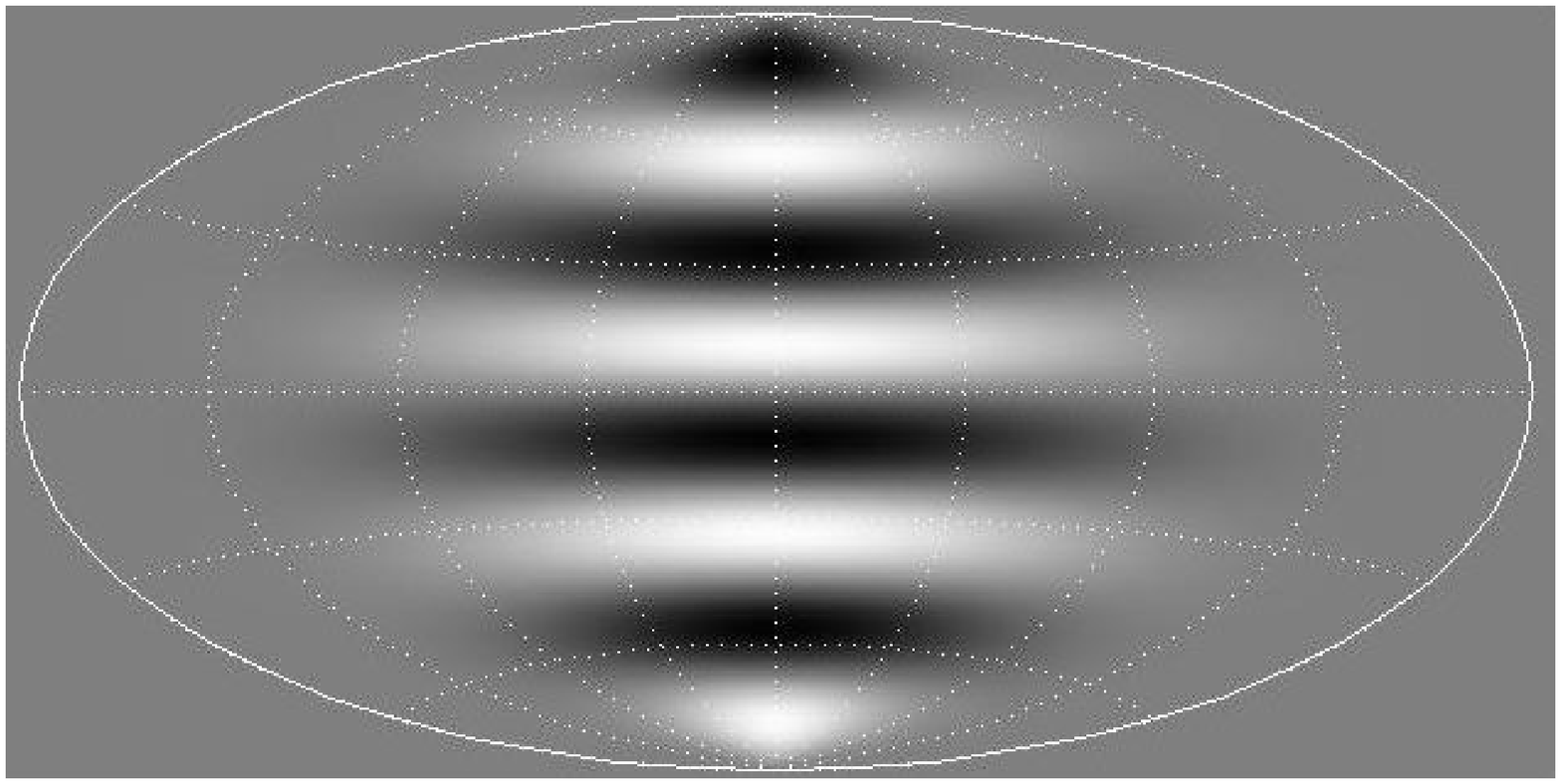,width=7cm}\epsfig{file=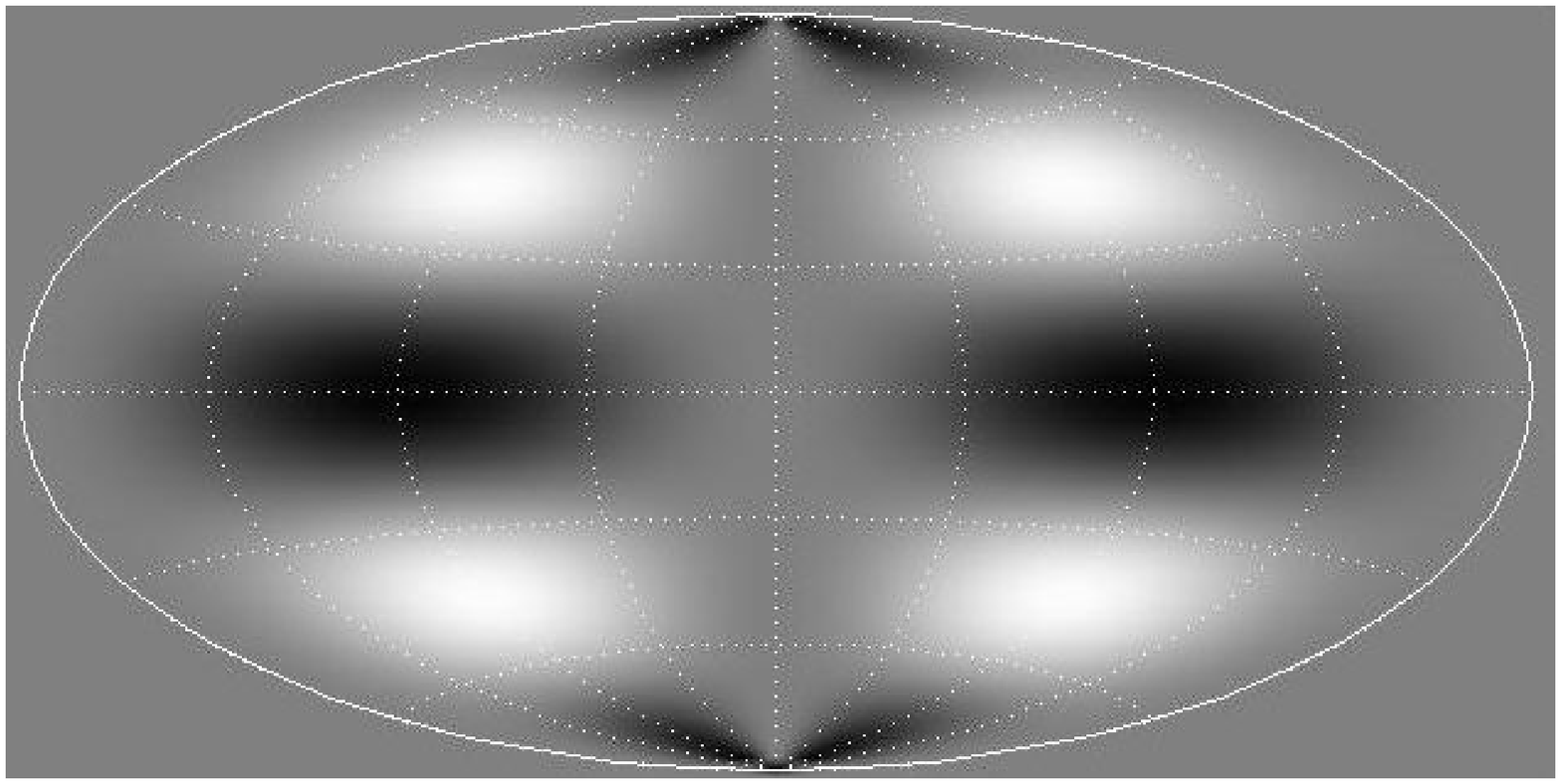,width=7cm}
\epsfig{file=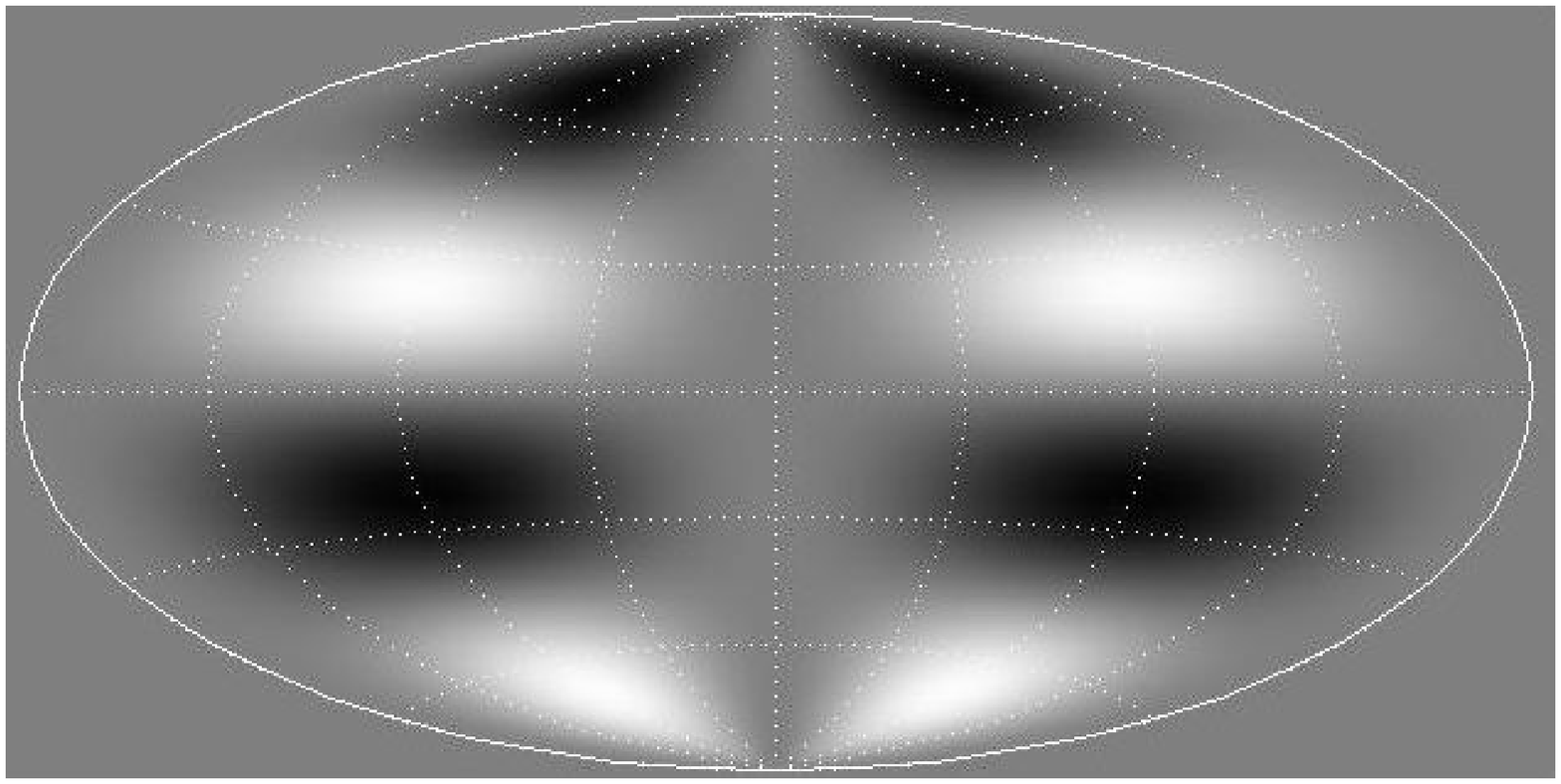,width=7cm}\epsfig{file=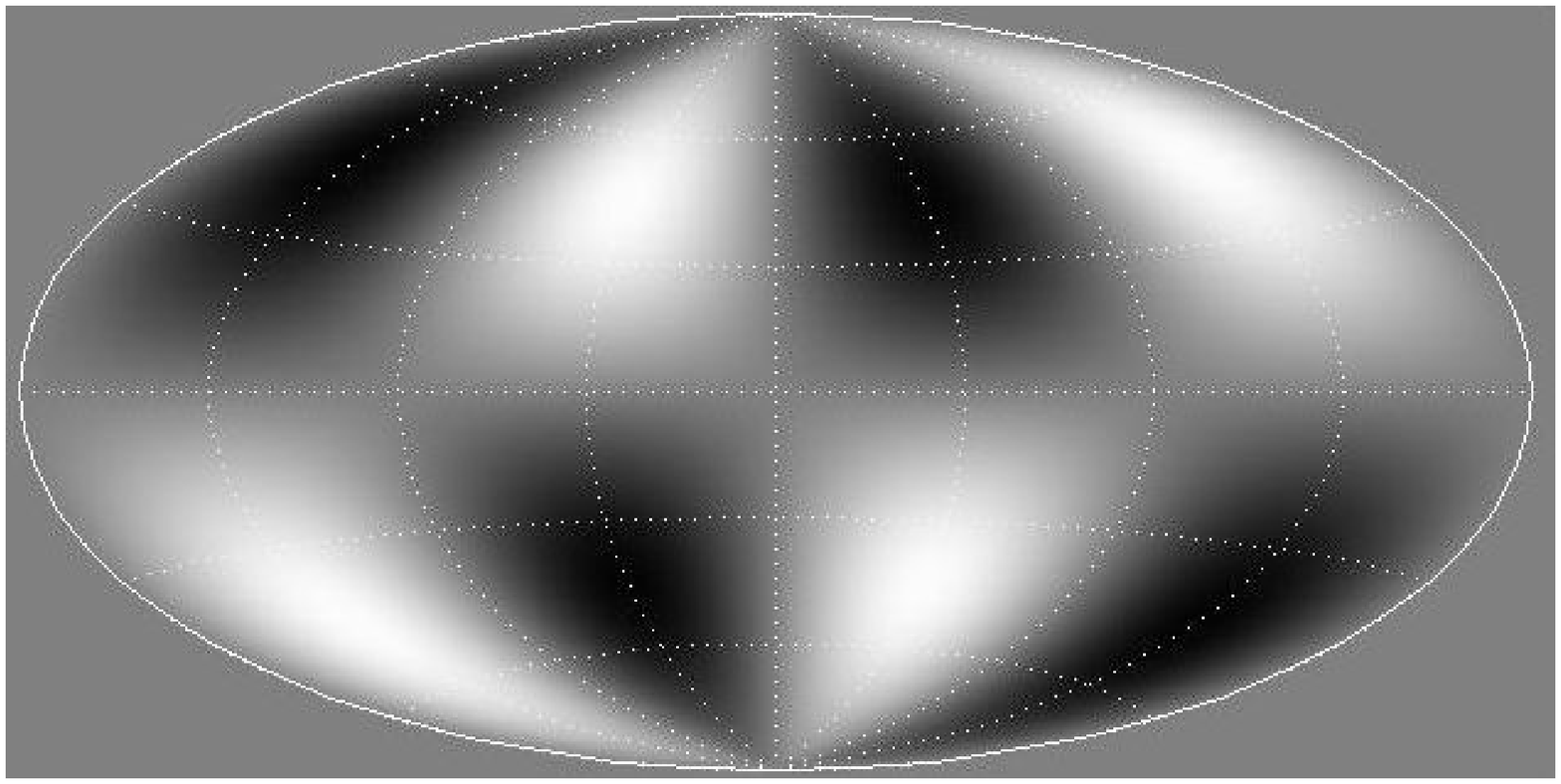,width=7cm}
\epsfig{file=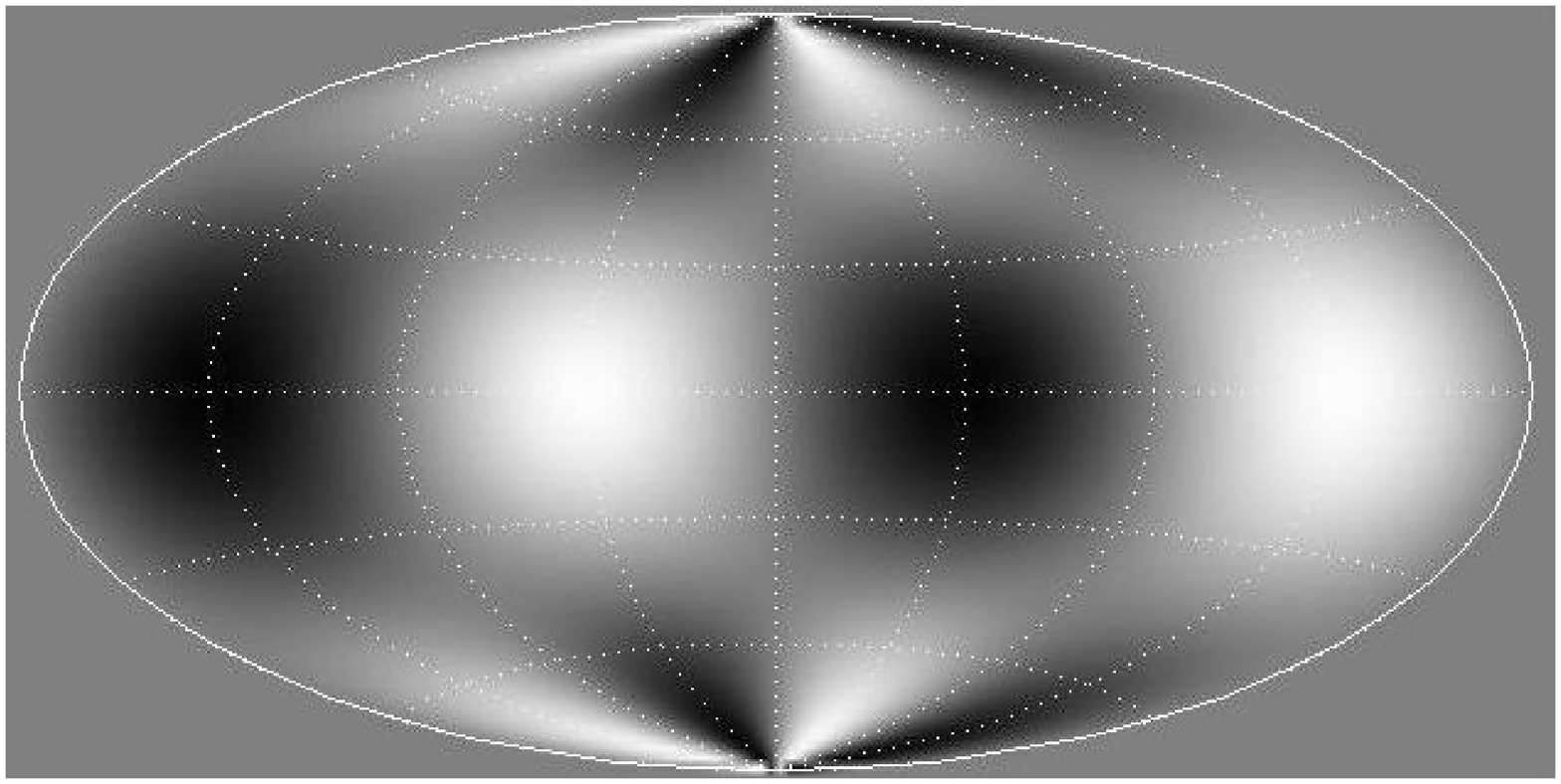,width=7cm}\epsfig{file=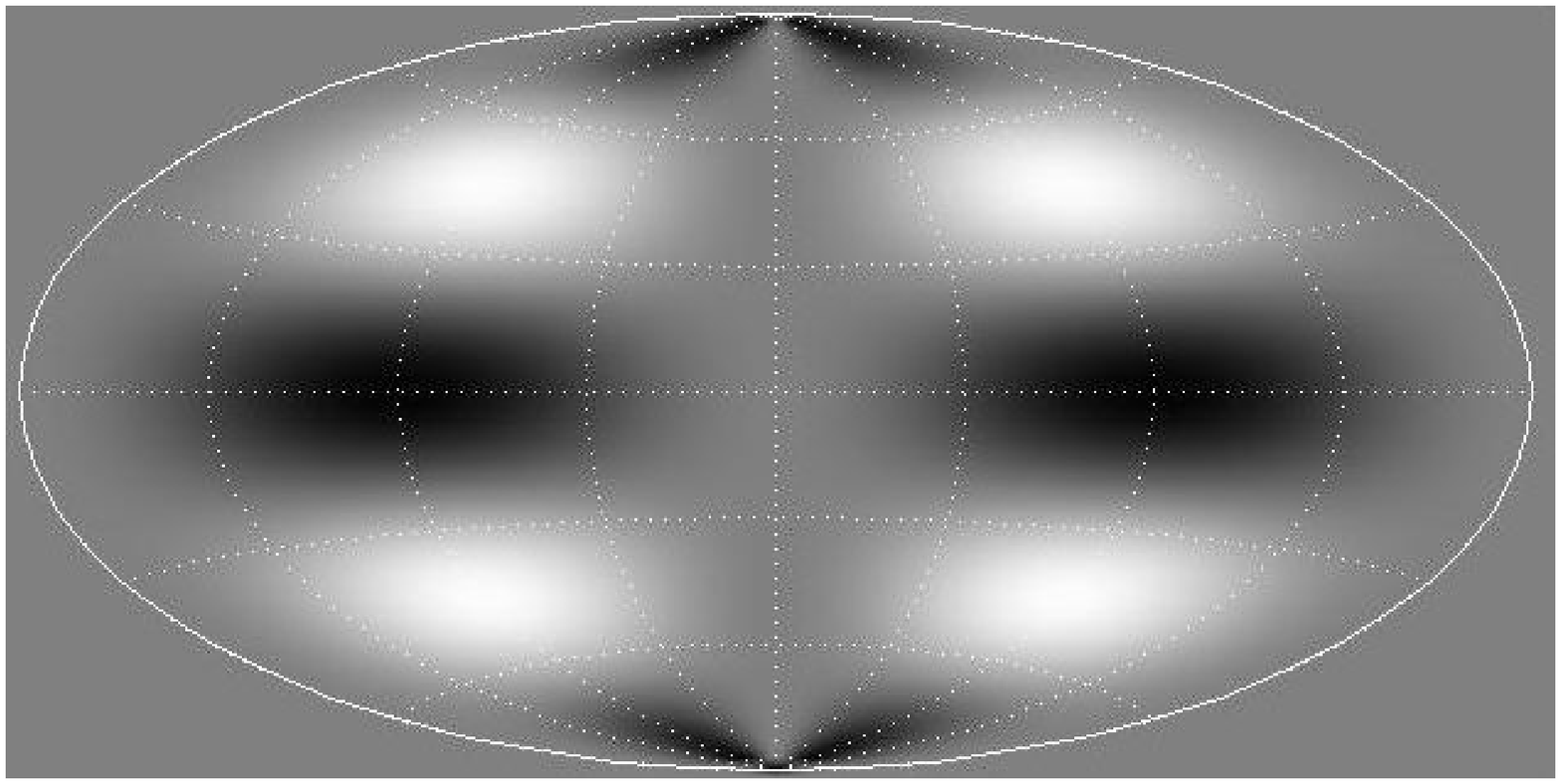,width=7cm}
\caption{The ten eigenmodes of $S^3/D_2^*$ for $k=4$ in Hammer-Aitoff
projection for $\chi=\pi/2$. Note that some are identical but only on
the equator and it can be checked that these modes are indeed linearly
independent.} \label{fig90}
\end{center}
\end{figure}

\begin{figure}
\begin{center}
\epsfig{file=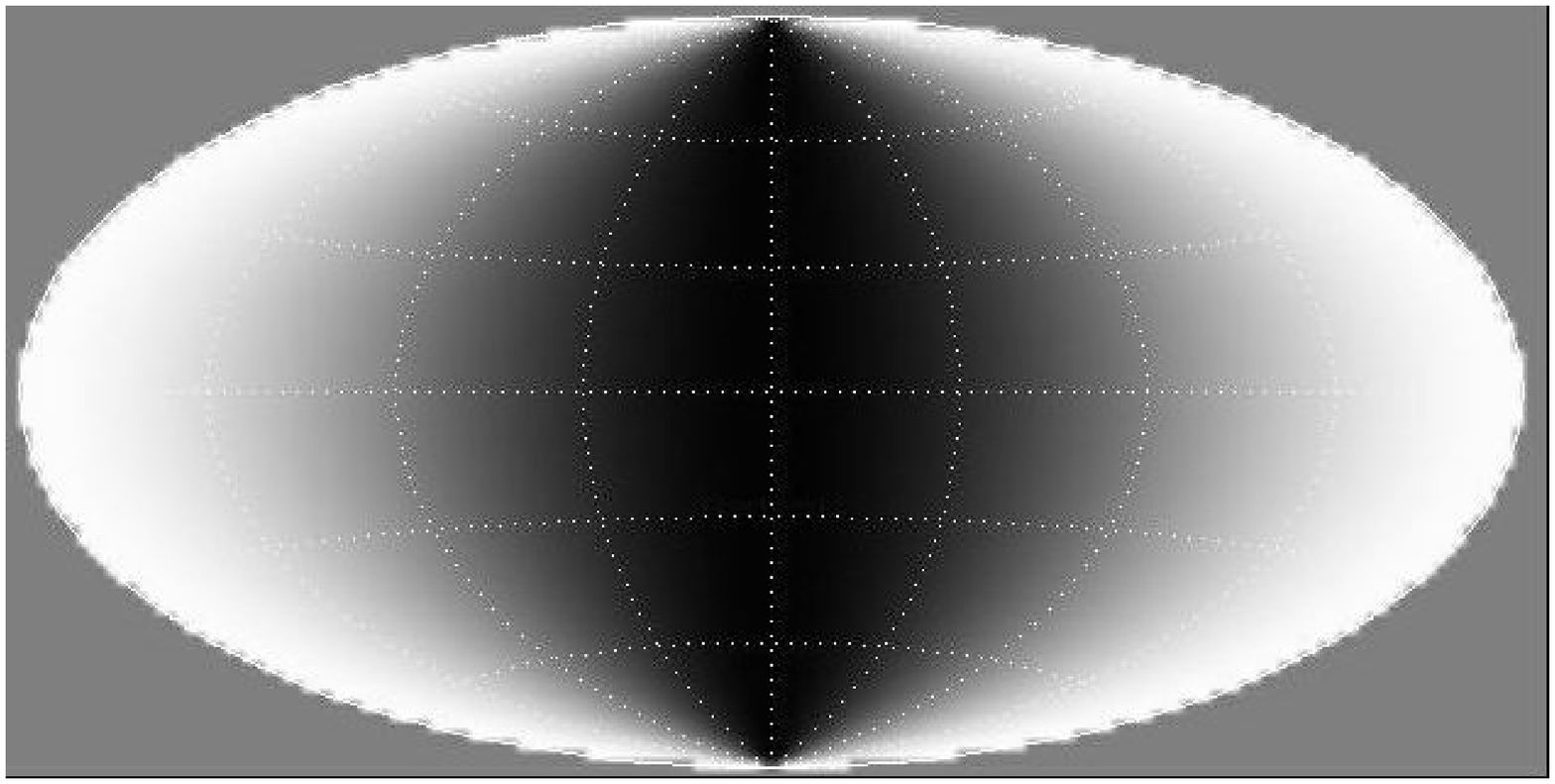,width=7cm}\epsfig{file=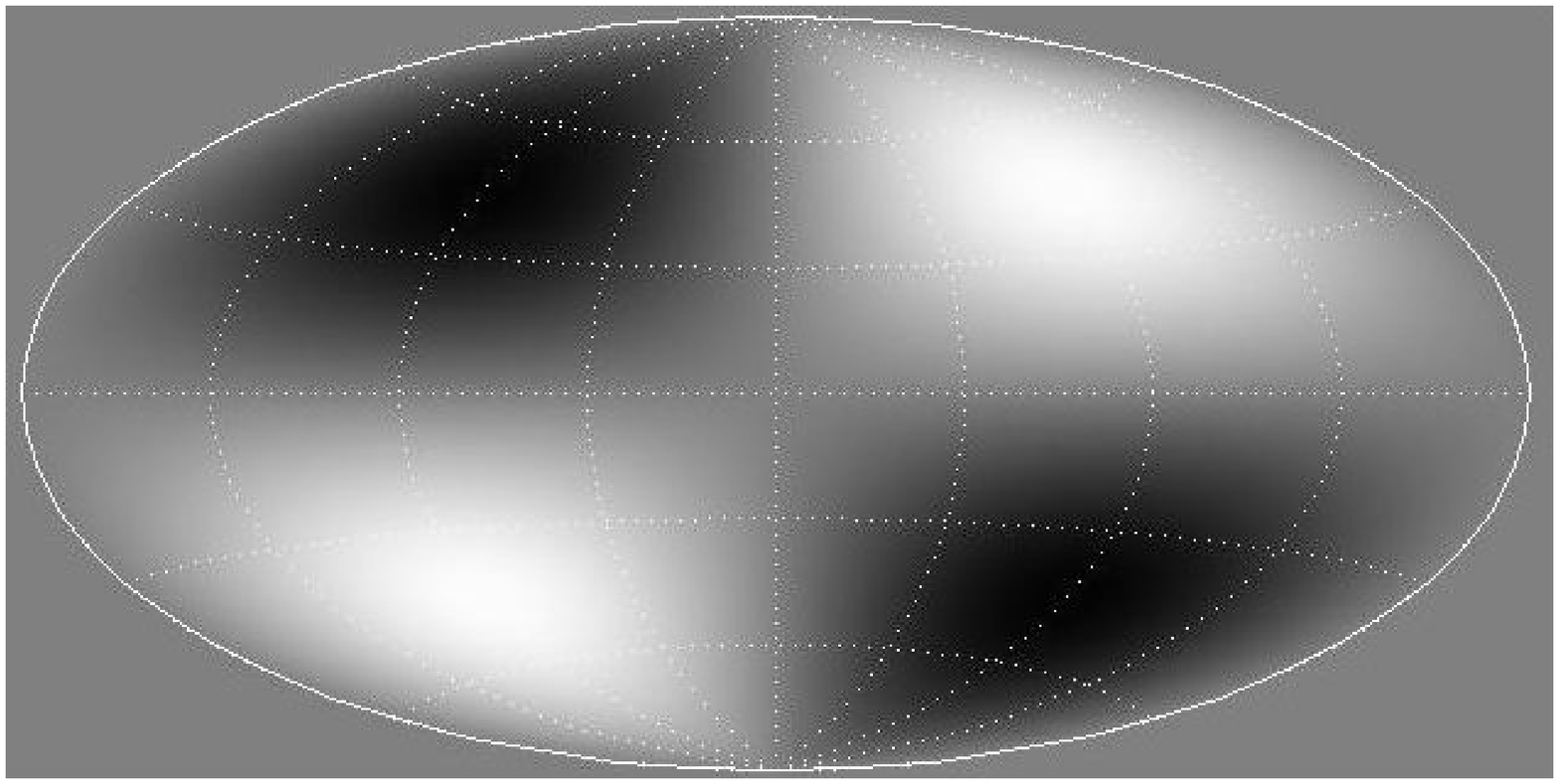,width=7cm}
\epsfig{file=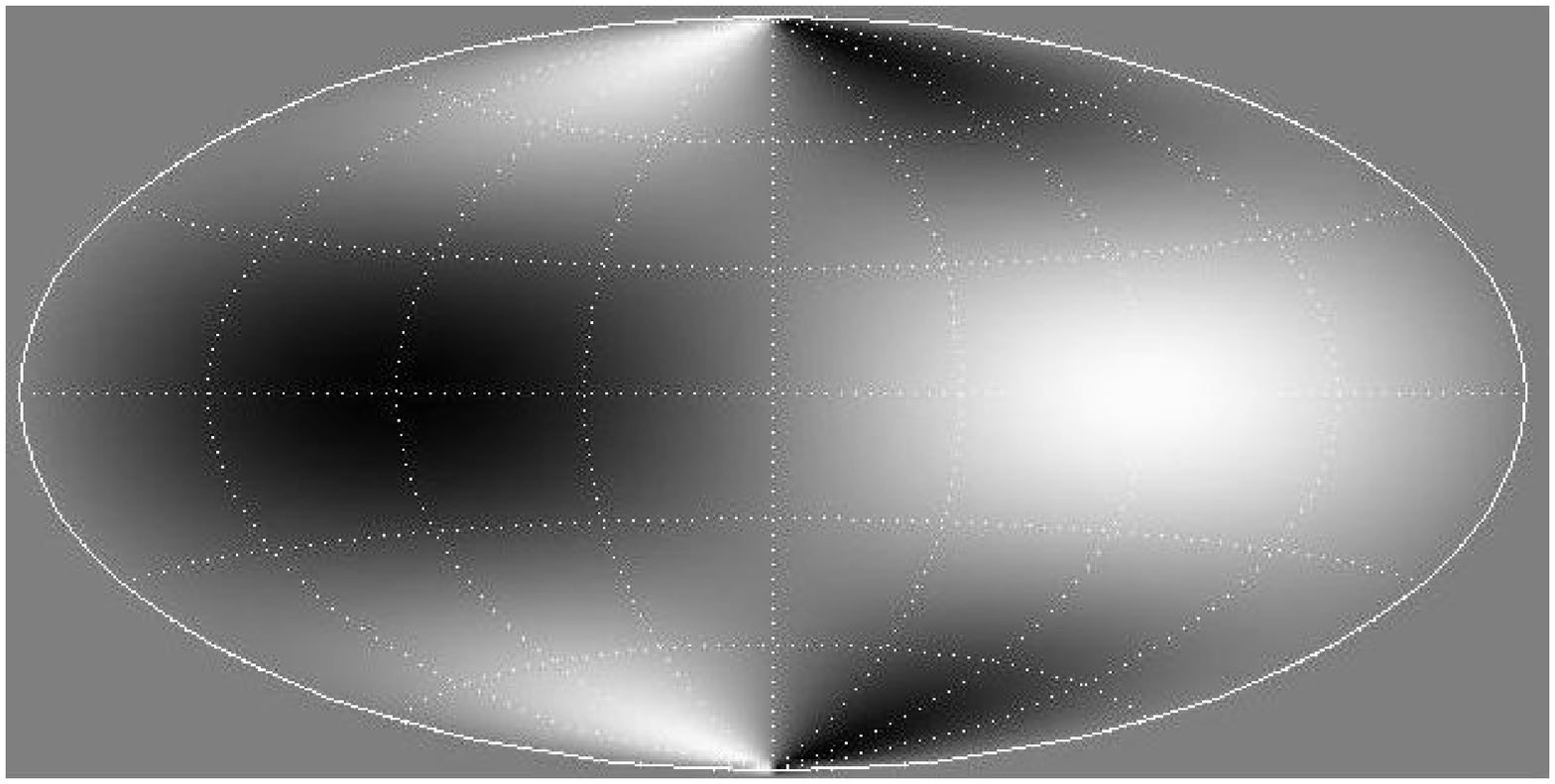,width=7cm}
\caption{The three eigenmodes of $S^3/Z_8$ for $k=2$ in Hammer-Aitoff
projection for $\chi=\pi/2$.} \label{fig92}
\end{center}
\end{figure}

\begin{figure}
\begin{center}
\epsfig{file=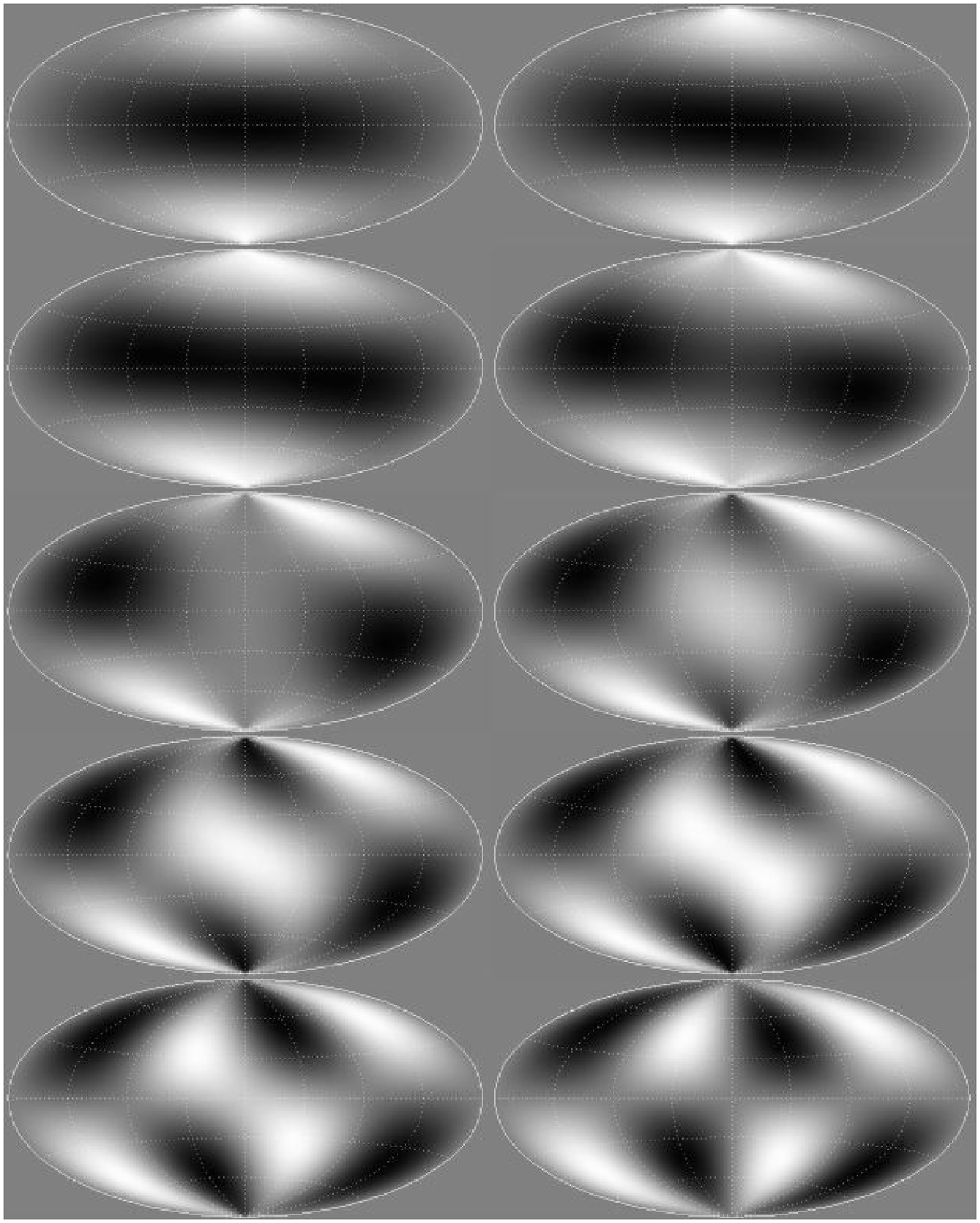,width=14cm}
\caption{A mode of $S3/D_2^*$ with $k=4$ for 10 different values of
$\chi=i\pi/20$ ($i=1..10$). It must be read left to right, top to
bottom.}
\label{fig93}
\end{center}
\end{figure}

In figures~\ref{fig90} and~\ref{fig92}, we present the two examples of
the lowest modes of $S^3/D_2^*$ and $S^3/Z_8$. In figure~\ref{fig93}
we depict one of the ten eigenmodes of $S^3/D_2^*$ with $k=4$ for
different values of the radial coordinate $\chi$.

\section{Some cosmological implications}
\label{sec_cosmo}

The goal of this section is not to compute the CMB anisotropies in
details (this task will be delt with in a follow-up
article~\cite{rglluw}), but to give estimate of the expected effects
on large angular scales.

The evolution of the scale factor, $a$, of the universe is dictated by
the Friedmann equation that can be recast under the form
\begin{equation}\label{fried1}
\left(\frac{{\cal H}}{{\cal H}_0}\right)^2=
\Omega_{r_0}x^{-2}+\Omega_{m_0}x^{-1}+
\Omega_{\Lambda_0}x^{2}+(1-\Omega_{r_0}-\Omega_{m_0}
-\Omega_{\Lambda_0})
\end{equation}
where ${\cal H}\equiv a'/a$, a prime denoting a derivative with
respect to the conformal time $\eta$. We have introduce $x\equiv
a/a_0\equiv(1+z)^{-1}$, $z$ being the redshift and $a_0$ the
value of the scale factor today. The density parameters are defined by
\begin{equation}
\Omega_{r_0}\equiv\frac{\kappa\rho_{r_0}}{3H_0^2},\quad
\Omega_{m_0}\equiv\frac{\kappa\rho_{m_0}}{3H_0^2},\quad
\Omega_{\Lambda_0}\equiv\frac{\kappa\rho_{\Lambda_0}}{3H_0^2}
\end{equation}
where $\kappa\equiv 8\pi G$ and $H_0={\cal H}_0/a_0$. In terms of
these quantities, the physical curvature radius today is given by
\begin{equation}\label{rco}
R_{C_0}^{\rm phys} \equiv a_{0}R_{C_0} =
\frac{c}{H_0}\frac{1}{\sqrt{\left|\Omega_{\Lambda_0}
 + \Omega_{m_0}-1\right|}}
\end{equation}
We can choose $a_0$ to be the physical curvature radius today, i.e.
$a_0 = R_{C_0}^{\rm phys}$, which amounts to choosing the units on
the
comoving sphere such that $R_{C_0} =1$, hence determining the value
of
the constant $a_0=c\left|\Omega_{\Lambda_0} +
\Omega_{m_0}-1\right|^{-1/2}/H_0$.

\subsection{Generalities}

When studying the CMB anisotropies, one has to go beyond the
homogeneous and isotropic description of our universe and need to
condider a perturbed spacetime with metric
\begin{equation}
\d s^2 = a^2(\eta)\left[-(1+2\Phi)\d\eta^2+
         (1-2\Psi)\gamma_{ij} \d x^i\d x^j \right]
\end{equation}
where we consider only scalar modes and working in longitudinal
gauge. Vector and tensor modes would have to be added for a
complete description (including for instance gravitational waves,
but are negligible on large angular scales.
On these scales, we can neglect the effect of the anisotropic
pressure so that $\Psi=\Phi$ and the effect of the radiation
between the last scattering surface and today.

Under these assumptions, the temperature fluctuation in a direction
${\bf n}$ can be related~\cite{sw,panek} to the gravitational
potential $\Phi$ by, again in the particular case of adiabatic initial
perturbations,
\begin{equation}\label{cosmo1}
\frac{\delta T}{T}({\bf n})=\frac{1}{3}\Phi[\eta_{_{\rm LSS}},
(\eta_0-\eta_{_{\rm LSS}}){\bf n}]+2\int_{\eta_{_{\rm
LSS}}}^{\eta_0} \frac{\partial\Phi[\eta,(\eta_0-\eta){\bf
n}]}{\partial\eta}\d\eta
\end{equation}
where $\eta_{_{\rm LSS}}$ and $\eta_0$ are the value of the conformal
cosmic time at the emission of the photon (last scattering surface)
and at the reception (observer). According to the standard
nomenclature, we will refer to the first term as the ordinary
Sachs-Wolfe term (OSW) and to the second as the integrated Sachs-Wolfe
term (ISW). The temperature angular correlation function is then
defined by
\begin{equation}
C(\theta)\equiv\left\langle\frac{\delta T}{T}({\bf n}_1)
\frac{\delta T}{T}({\bf n}_2)\right\rangle_{\cos\theta={\bf
n}_1.{\bf n}_2}.
\end{equation}
Decomposing the temperature fluctuation on the spherical harmonics
as
\begin{equation}\label{dev}
\frac{\delta T}{T}({\bf n})=\sum_\ell\sum_{m=-\ell}^\ell a_{\ell
m}Y_{\ell
m}(\theta,\phi)
\end{equation}
the coefficients $C_\ell$ of the development of $C(\theta)$ on
Legendre polynomials are given by
\begin{equation}
(2\ell+1)C_\ell=\sum_{m=-\ell}^\ell\left\langle a_{\ell m}a_{\ell
m}^*\right\rangle.
\end{equation}

To compute these quantities, one needs to determine the gravitational
potential $\Phi$. Its evolution is dictated~\cite{kodama} by the
equation
\begin{equation}\label{evo}
\Phi''+3{\cal H}(1+c_s^2)\Phi'-c_s^2\Delta^2\Phi +[2{\cal
H}'+(1+3c_s^2)({\cal H}^2 -K)]\Phi=0.
\end{equation}
This relation strictly holds only for initial adiabatic
perturbations as predicted by most of the inflationary scenarios.
$c_s^2=P'/\rho'$ is the sound speed and is given by
$c_s^2=\rho_r/(\rho_m+4/3\rho_r)/3$. After decomposing $\Phi$ on
the eigenmodes as
\begin{equation}\label{p1}
  \Phi(\eta,\bx)=\sum_{\beta,s}\Phi_{\beta,s}\Psi_{\beta,s}^{^{[\Gamma]}}(\bx),
\end{equation}
one can easily show that if the universe is matter dominated
between the last scattering epoch and today then
\begin{equation}\label{p2}
 \Phi_{\beta,s}(\eta)=F(\eta)\Phi_0(\bq)
\end{equation}
where we use the notation that $\bq=(\beta,s)$.  $\Phi_0$ is the value
of the gravitational potential at the beginning of the matter era, but
since in the early universe the curvature term is negligible and the
dynamics is dominated by the radiation (so that $a\propto\eta$) it can
be shown that, for long wavelengths ($k\eta\ll1$), the non decaying
mode of the gravitational potential is constant so that $\Phi_0$ is in
fact the primordial gravitational potential.  Inflationary theories
predict that it is a Gaussian field, and that all modes are
independent, with power spectrum~\cite{ks}
\begin{equation}\label{p3}
\left\langle\Phi_0(\bq)\Phi_0^*(\bq')\right\rangle=
\frac{2\pi^2}{\beta(\beta^2-K)}
{\cal P}_\Phi(\beta)\delta(\beta-\beta')\delta_{ss'}.
\end{equation}
where $K = 0, \pm 1$ is the sign of the curvature.  In the case of a
scale invariant Harrison-Zel'dovich spectrum, ${\cal
P}_\Phi\propto\beta^0$.

Now, inserting the decomposition (\ref{p1}) with the solution
(\ref{p2}) in Eq.~(\ref{cosmo1}) and decomposing the eigenmodes as
in Eq.~(\ref{c3}), one finally gets that the coefficients of the
development (\ref{dev}) are given by
\begin{equation}\label{p4}
  a_{\ell m}=\sum_{\beta,s}\Phi_0(\beta,s)\xi_{\beta,s\ell m}
  G_{\beta\ell}
\end{equation}
where $G_{\beta\ell}$ is defined by
\begin{equation}\label{p5}
G_{\beta\ell}\equiv\frac{1}{3}F(\eta_{_{\rm
LSS}})R_{\beta\ell}(\eta_0-\eta_{_{\rm
LSS}})+2\int_{\eta_0}^{\eta_{_{\rm
LSS}}}F'(\eta)R_{\beta\ell}(\eta_0-\eta)\d\eta.
\end{equation}
The correlation
\begin{equation}\label{co}
\left<a_{\ell
m}a_{\ell'm'}\right>=\sum_{\beta,s}\frac{2\pi^2}{\beta(\beta^2-K)}
{\cal P}_\Phi(\beta)G_{\beta\ell}G_{\beta\ell'}\xi_{\beta,s\ell m}
\xi_{\beta,s\ell'm'}^*
\end{equation}
has non-zero off-diagonal terms which reflects the fact that there is
a global anisotropy due to the non--trivial topology. In a simply--connected
homogeneous and isotropic universe $\left<a_{\ell
m}a_{\ell'm'}\right>=C_\ell\delta_{\ell\ell'}\delta_{mm'}$. These
off-diagonal terms are characteristic of the non-Gaussianity induced
by the topology. From the expression (\ref{co}), we can extract the
$C_\ell$ which characterise only the isotropic part of the
temperature distribution, as
\begin{equation}\label{p6}
(2\ell+1)C_\ell^{^{[\Gamma]}}=\sum_{\beta,s,
m}\frac{2\pi^2}{\beta(\beta^2-K)}{\cal
P}_\Phi(\beta)\left|\xi_{\beta,s\ell m}\right|^2
\left|G_{\beta\ell}\right|^2.
\end{equation}
Indeed, in the Euclidean and hyperbolic cases, the sum over $\beta$
has to be replaced by an integral; in the spherical case, $\beta\geq
{\rm max}(3,\ell+1)$.  This result has to be compared to its covering
space analog, obtained by setting $\xi_{\beta,s\ell m}=1$ when the
index $s$ can be assigned the value $(\ell m)$ and zero otherwise, so
that the sum over $m$ gives $(2\ell+1)$,
\begin{equation}\label{p7}
C_\ell^{^{[U]}}=\sum_{\beta}\frac{2\pi^2}{\beta(\beta^2-K)}{\cal
P}_\Phi(\beta) \left|G_{\beta\ell}\right|^2,
\end{equation}
so that there is an average effect with the ponderation
$\left|\xi_{\beta,s\ell m}\right|^2/(2\ell+1)$.

The effect of the topology on large scales can thus be
investigated, in a first step, by considering the index
\begin{equation}
\Upsilon_\ell^{^{[\Gamma]}}\equiv\left\vert\Gamma\right\vert
\frac{C_\ell^{^{[\Gamma]}}}{C_\ell^{^{[U]}}}.
\end{equation}
At large angular scales, $\Upsilon_\ell$ will oscillate due to the
missing eigenmodes while it will converge to unity, mainly because
of Eq.~(\ref{weyl3}), on small angular scales where the topology
becomes irrelevant. Indeed, this will allow to put constraints on
some topologies but an unambiguous detection will have to use a
full-sky CMB map.

\subsection{Example of the three-torus}

As a first example, let us consider a cubic 3-torus of comoving size
$L$. In the simplest case in which $\Omega_\Lambda=0$, $F$ is constant
and
\begin{equation}\label{p8}
C_\ell^{^{[T^3]}}=\sum_{\bk}\frac{{\cal
P}_\Phi(k)}{k^3L^3}j_\ell^2(k(\eta_0-\eta_{_{\rm LSS}})),
\end{equation}
with $\bk=2\pi(n_1,n_2,n_3)/L$, and for the universal covering space
we have
\begin{equation}\label{p9}
C_\ell^{^{[U]}}\propto\int\frac{\d
u}{u}j_\ell^2(u)\propto\frac{1}{\ell(\ell+1)}
\end{equation}
known as the Sachs-Wolfe plateau.

On this simple example, we can determine the cut-off $\ell_{\rm cut}$
below which there is a suppression of the power spectrum. The first
approach is to remind that the Bessel functions peak at
$k(\eta_0-\eta_{_{\rm LSS}})\sim\ell$. Hence a mode $k$ contributes
maximally to the angle subtended by its corresponding scale at last
scattering. In flat models, $\eta_0$ is given~\cite{peebles} by
\begin{equation}
\eta_0\simeq\frac{2ca_0}{H_0\sqrt{\Omega_{m_0}}}\left(1+\ln
\Omega_{m_0}^{0.085}\right).
\end{equation}
It follows that the OSW term has a cut-off round
\begin{equation}\label{i9}
\ell_{\rm cut}\sim\frac{4\pi c}{L_0H_0\sqrt{\Omega_{m_0}}}
\left(1+\ln\Omega_{m_0}^{0.085}\right).
\end{equation}
This is analogous to the estimate by Inoue~\cite{inoue01}.

Another approach, first introduced in~\cite{jpu1,jpu2}, is to compute
the angle $\theta_{\rm cut}$ under which the maximum comoving
wavelength $\lambda_{\rm max}=L$ at last scattering. It is given
by
\begin{equation}
\theta_{\rm cut}=\frac{a_0L}{(1+z)\d_{\rm A}(z)}
\end{equation}
where $\d_{\rm A}=\d_{\rm L}/(1+z)^2$ is the angular distance and
where the luminosity distance is given by
\begin{equation}
\d_{\rm L}(z)=(1+z)\frac{ca_0}{H_0}f\left[\int_0^z\frac{\d
u}{(1+u)E(u)} \right].
\end{equation}
In the case where
$\Omega_\Lambda=0$, $\d_{\rm L}(z)=2a_0(1+z)(1-(1+z)^{-1/2})/H_0$ so that
$\theta_{\rm cut}\sim H_0L/2$.  We have introduced $E\equiv{\cal
H}/{\cal H}_0$.  Now, the $\ell^{\rm th}$ Legendre polynomial is a
polynomial of degree $\ell$ in $\cos\theta$ and has $\ell$
zeros in $[-1,1]$ (or $2\ell$ zeros in $[-\pi,\pi]$ if working in
$\theta$) with approximatively the same spacing.  We can estimate
that
$\theta\sim\pi/\ell$ and thus that the cut in the OSW contribution is
expected to be round
\begin{equation}
\ell_{\rm cut} \sim \frac{2\pi c}{H_0L}f\left[\int_0^{z_{_{\rm
LSS}}}\frac{\d u}{(1+u)E(u)} \right],
\end{equation}
the factor 2 arising from the fact that an oscillation corresponds to
2 zero. This corresponds exactly to the previous estimate (\ref{i9}).

\subsection{Spherical spaces}

In the spherical case we expect the same kind of effects, i.e. a
suppression of the large scale ISW effect due of the existence of a
maximal wavelength. The OSW term will be approximatively the same as
the one computed in the flat case since our universe, even if closed,
is still very flat.

Let us first remark a crucial difference between Euclidean manifolds
and spherical manifolds. For the former, as we have seen in the
previous section, the smallest multipole is directly related to the
size $L$ of the fundamental polyhedron so that one cannot consider too
small universes. For spherical universes, the situation is {\it a
priori} different. As seen on the example of single-action manifolds,
the value of the first non-zero eigenvalue does not depend on the
order of the group, at least for cyclic and binary dihedral
groups.

Let us take the example of lens spaces, the first nonzero eigenvalue
is indeed the same for all $S^3/Z_m$ ($m > 1$), namely $k = 2$ and
eigenvalue 2(2+2) = 8.  However, the multiplicity is 3 for homogeneous
lens spaces $L(p,1)$, but only 1 for nonhomogeneous lens spaces
$L(p,q)$, as shown in Table 1 of~\cite{wlu02}. This can be understood
by the fact that the space is becoming smaller and smaller in {\it
only one} direction.  In perpendicular directions the space remains
large.  The waves do not have distinct peaks (like mountaintops found
in nature) but rather have ridges (perfect horizontal ridges, which
are never found in nature).  First imagine a wave in $S^3$, as
follows: the set of maxima is a great circle, while the set of minima
is a complementary great circle.  As time passes, the wave goes up and
down, so what is the top at a time $t = 0$ becomes the bottom at time
$t = 1/2$ say, and vice versa, returning to its original position at
time $t = 1$.  Midway between the ``top ridge" and the ``bottom ridge"
is a torus which remains at height 0 for all times $t$.  In toroidal
coordinates (\ref{CoordinateDefinition}), one ridge is the circle at
$\chi = 0$, the other ridge is the circle at $\chi' =\pi/2$, and fixed
torus lies at $\chi' =\pi/4$. This wave is preserved by {\it all}
corkscrew motions along the natural axes, so it projects down to an
eigenmode of {\it all} lens spaces $L(p,q)$. In the notations of
Section~\ref{sec_torus}, it is given in toroidal coordinates as
$$\Psi_{200}(\chi', \theta', \phi') = \sin^2 \chi' - \cos^2 \chi' = - \cos
2\chi'$$
thus verifying that it is a wave as described above, and
in particular with no dependence on $\theta'$ or $\phi'$.
In the special case of a homogeneous lens space $L(p,1)$,
we get two more eigenmodes
$$\sin(2 \chi') \cos(\theta' - \phi')\qquad{\rm and}\qquad
\sin(2\chi')\sin(\theta' - \phi').$$
These modes are constant along helices (where $\theta' - \phi'$ is
constant), so they are eigenmodes of all homogeneous lens spaces,
whose holonomies are Clifford translations, but not eigenmodes of
nonhomogeneous lens spaces.

We thus expect the effect of the cut-off in the Sachs-Wolfe plateau to
be milder than for Euclidean spaces. But, on the other hand we expect
to have a more irregular Sachs-Wolfe plateau due of the fact that some
wavelengths are missing from the spectrum. Another observational
consequences missed by the angular power spectrum is a global large
scale anisotropy that, at least, is expected for non-homogeneous
lens spaces.

To illustrate this, consider the simplest example $S^3/Z_2$ for which
half of the modes have disappeared so that $k=2p$, $p$ being an
integer. It follows that for $\ell\sim2p$, $C_\ell^{^{[Z_2]}}\sim
C_\ell^{^{[S^3]}}$ and for $\ell\sim(2p+1)$,
$C_\ell^{^{[Z_2]}}\sim0$. We thus expect that the $C_\ell$ curve
oscillates around $C_\ell^{^{[S^3]}}/2$ with a frequency of order
$\ell\sim\eta_0$.

\begin{figure}
\begin{center}
\epsfig{file=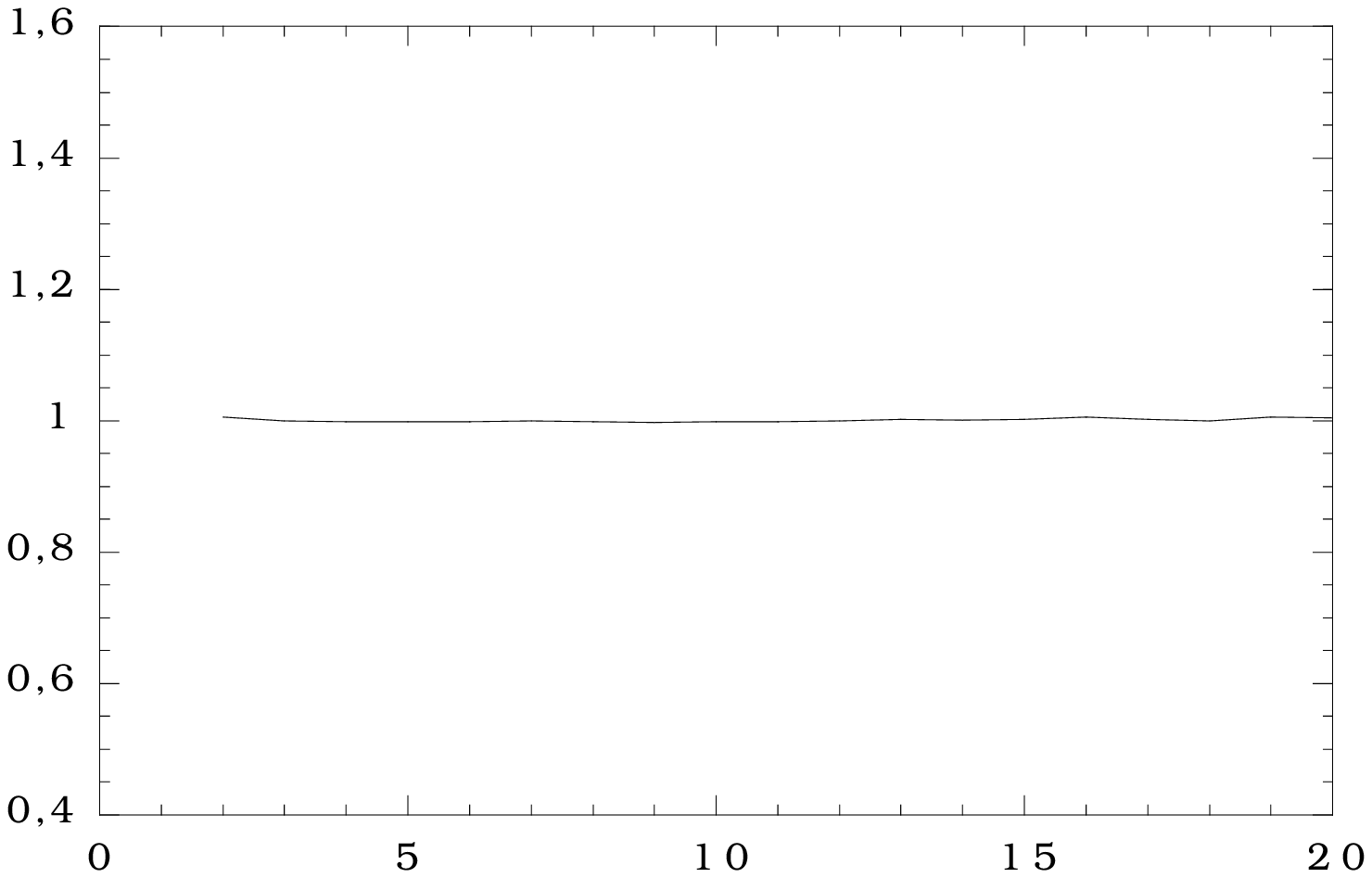,width=8cm}\epsfig{file=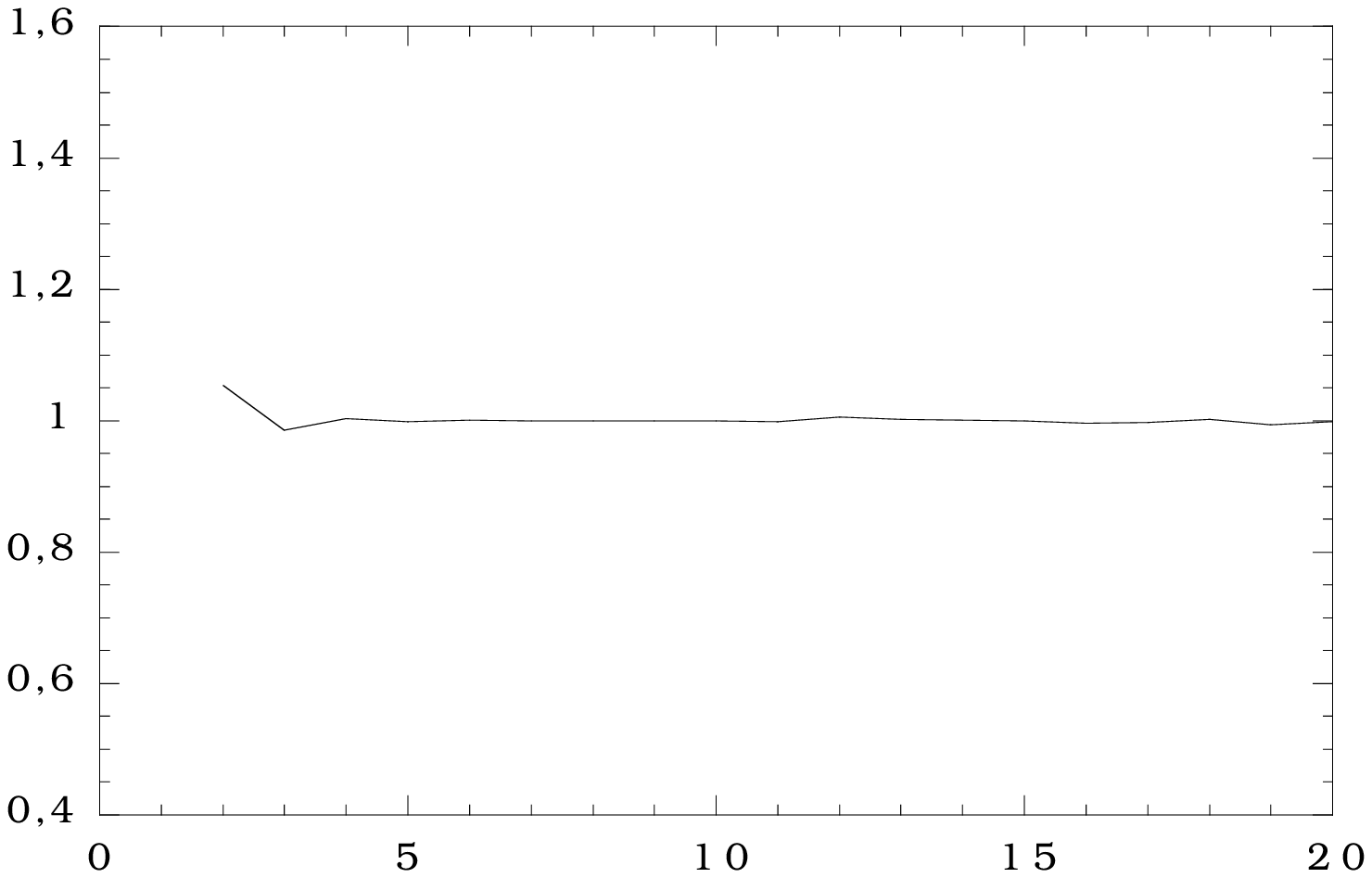,width=8cm}
\epsfig{file=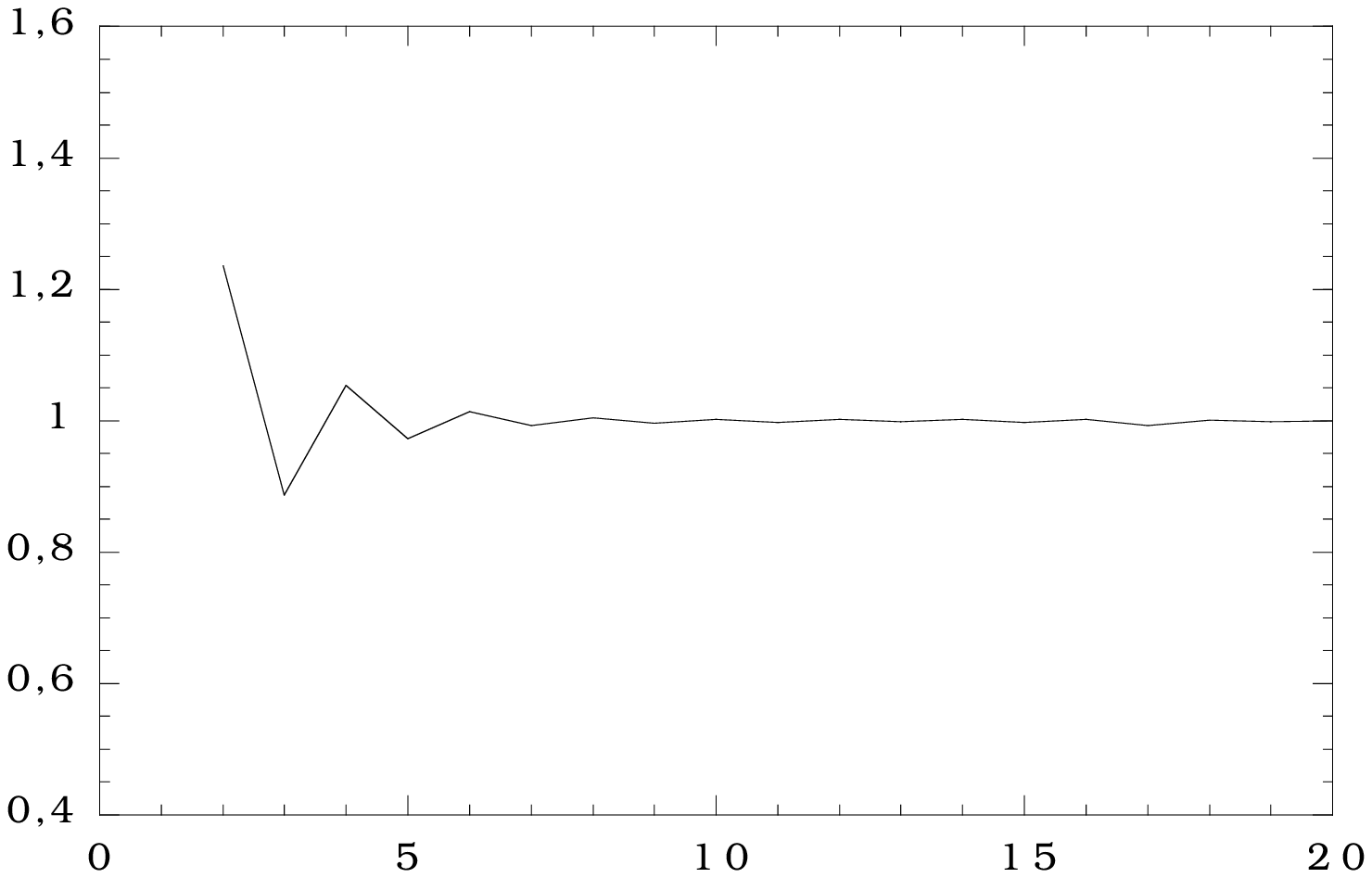,width=8cm}\epsfig{file=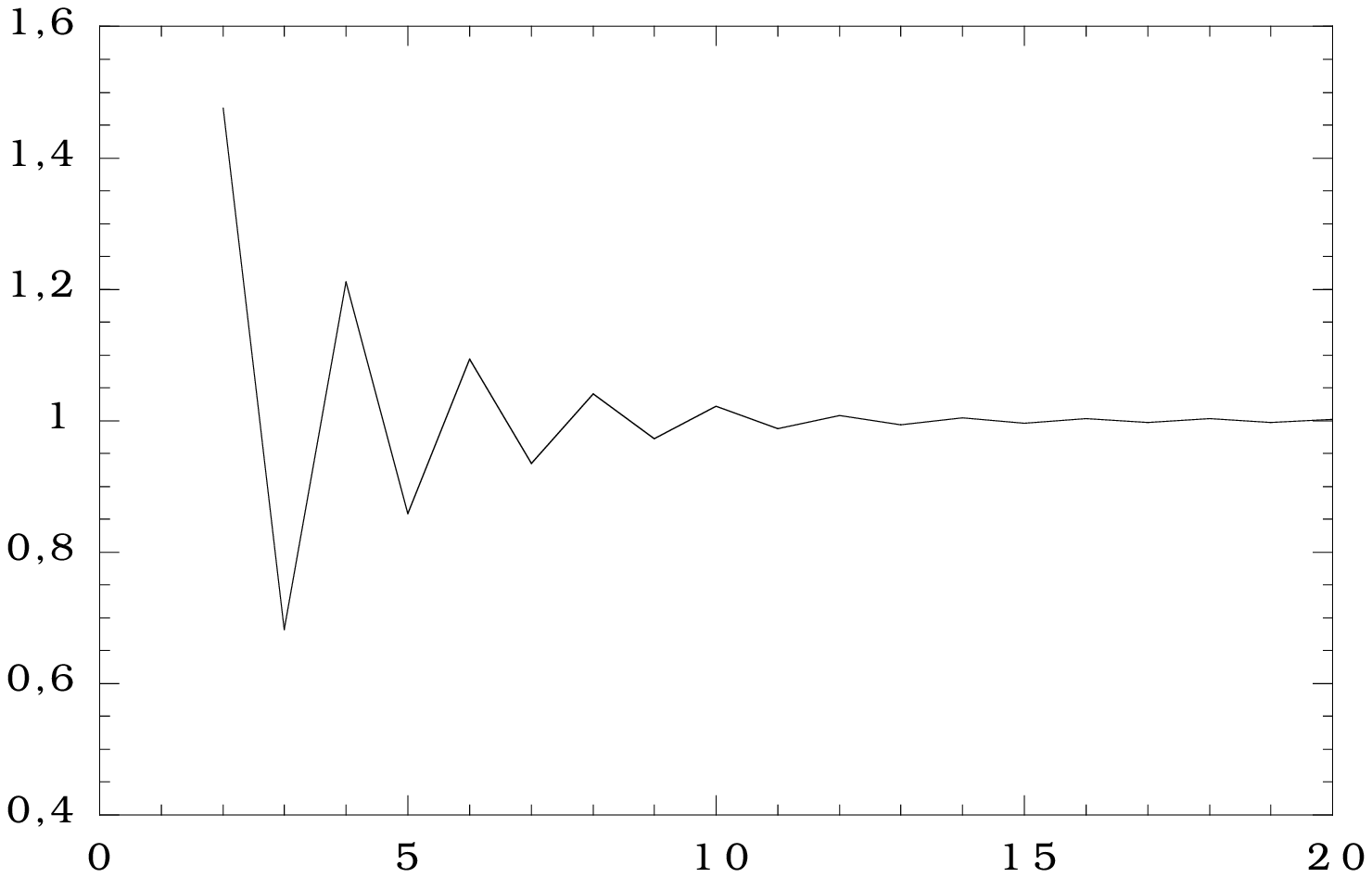,width=8cm}
\epsfig{file=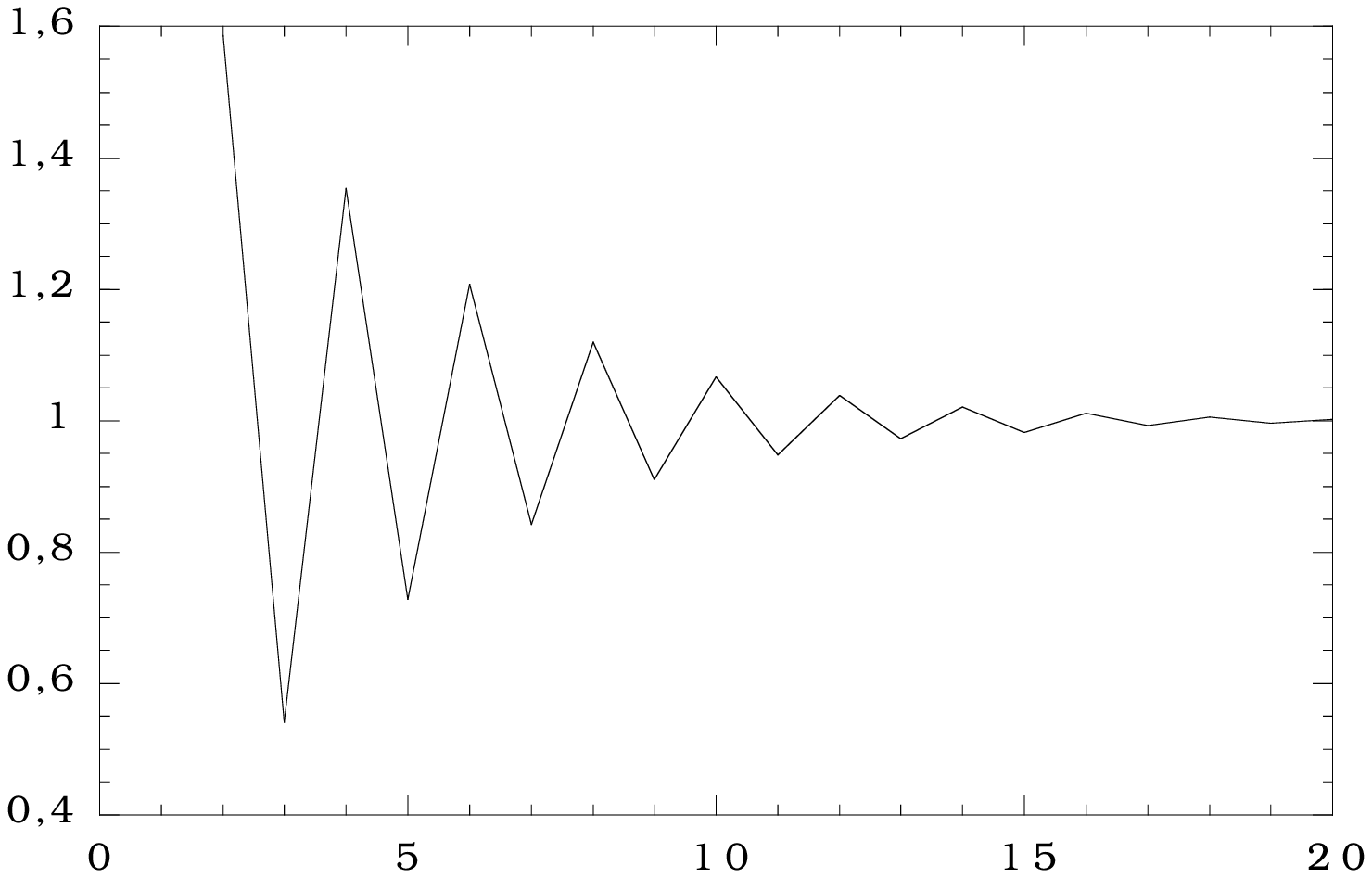,width=8cm}\epsfig{file=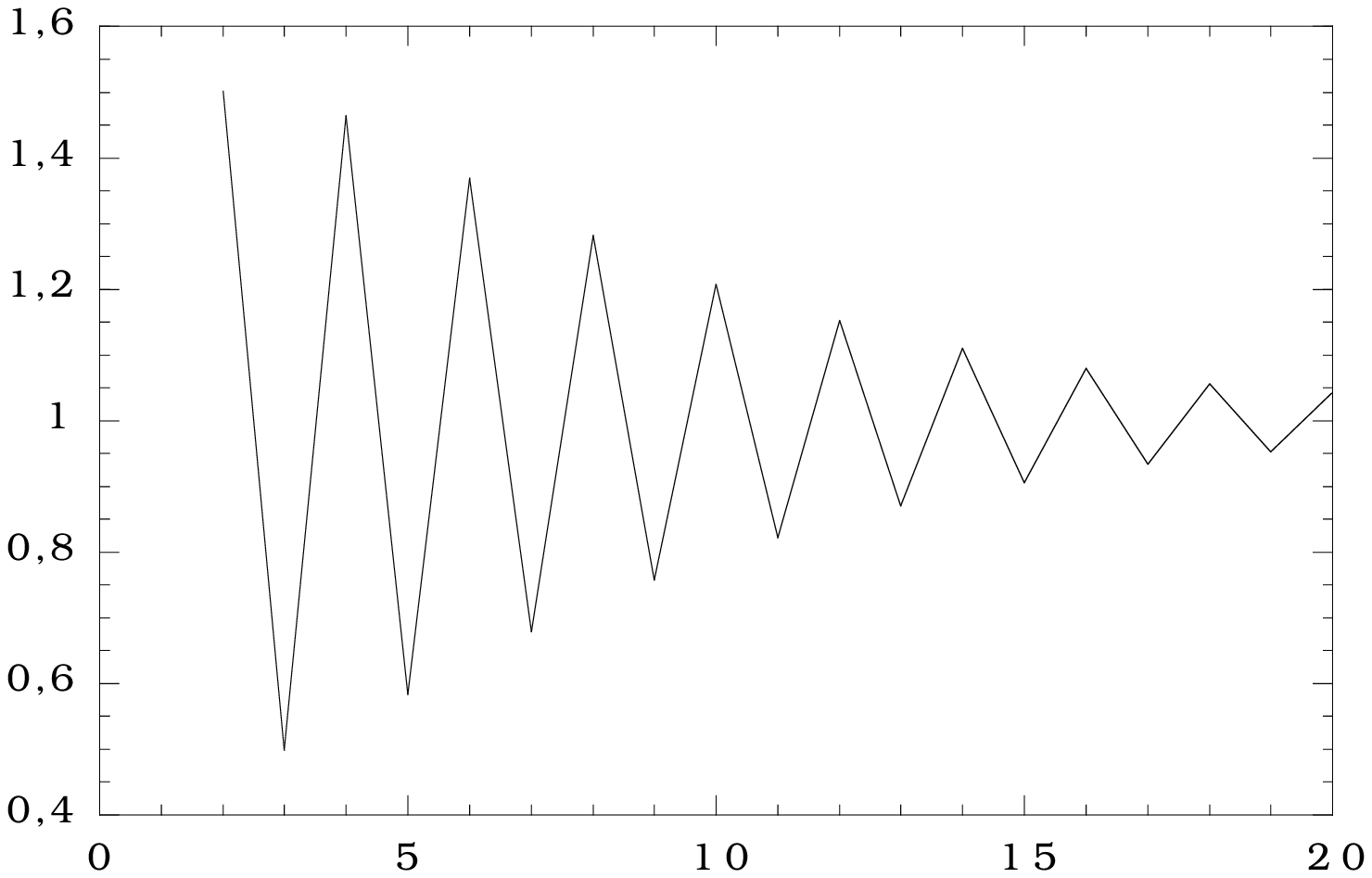,width=8cm}
\caption{The $\Upsilon_\ell$ for $\ell<20$ for $S^3/Z_2$. From top to
bottom and left to right the curvature increases as
$\Omega=1.1,1.25,1.5,1.7,1.8,1.9$ with $\Lambda=0$. This corresponds
respectively to $\chi_{_{\rm LSS}}=
0.5939,0.8990,1.1945,1.3528,1.4173,1.4746$. For higher curvature, the
topologicaal scale has more and more importance. The growth of the
amplitude of the oscillations at small multipole is dur to the growth
of the $C_\ell$ due to the integrated Sachs-Wolfe effect.}
\label{fig9}
\end{center}
\end{figure}

Using the coefficients (\ref{5}) for the projective space, we plot in
figure~\ref{fig9}, the function $\Upsilon_\ell$ for $\ell\leq20$ for
different curvature radius. We do not include the cosmic
variance. This confirms the previous semi-analytical analysis and is
in agreement with the numerical computations performed for
$S^3/Z_2$~\cite{souradeep}. When $\vert\Omega-1\vert\ll1$,
$C_\ell^{^{[Z_2]}}\sim C_\ell^{^{[S^3]}}$ because $\chi_{_{\rm
LSS}}<\pi/2$, the topological scale. When $\chi_{_{\rm
LSS}}\sim\pi/2$, both antipodal points on the last scattering surface
are close to the equator so that the angular correlation function in
opposite directions is expected to be higher. The case in which
$\chi_{_{\rm LSS}}>\pi/2$ is even more intricate because the geodesics
are warping around the universe more than once.

Such features are general to all spherical spaces, but the higher
the order of the group, the larger the minimal curvature radius to
get a topological signal. Since the size of the manifold decreases
with the order of the group, there will always exist potentially
detectable topologies even for spaces very close to flatness.

Note also that in the case of the projective space,
$$
\left<a_{\ell m}a_{\ell'm'}\right>\propto\delta_{\ell\ell'}\delta_{mm'}
$$
which can be understood if one remembers that there is no breaking
of global isotropy and homogeneity for the projective plane. This
is the only exception for which one can have a non-trivial
topology and no preferred direction.

\section{Conclusion}
\label{sec_concl}

In this article, we have investigated in details the structure of the
eigenmodes of the Laplacian operator in spherical spaces. A series of
analytical and mathematical results have been either reviewed or
obtained and we have introduced various efficient numerical methods to
compute them. These methods were compared together and to some
analytical results.

We have also investigated some cosmological consequences of this work
par\-ti\-cu\-lar\-ly concerning the large angular scales of the cosmic
microwave background anisotropies. The effect of the topology in CMB
calculation has been described and these results are now being
included to a full Boltzmann code to simulate CMB maps with a
topological signal~\cite{rglluw}. As an example, we considered the
simplest of all cases, that is the projective space.

\appendix

\section{Eigenmodes of the Laplacian of homogeneous and isotropic
simply
connected three-dimensional spaces}\label{appA}

This appendix follows the work by Abbott and Schaeffer~\cite{abott86}
and Harrison~\cite{harrison67}.  Its goal is to summarize the
derivation and explicit forms of the scalar harmonic functions
solutions of the Helmoltz equations (\ref{Helmoltz1}).

We rewrite the Friedmann metric (\ref{fl_metric}) as
\begin{equation}\label{eq:IsotropicMetric}
    \d s^2 = -\d t^2 + \frac{a^2(t)}{\sigma^2(r)}
    \left(\d r^2 + r^2 \d\Omega^2\right)
\end{equation}
with $\sigma(r) = 1 + Kr^2/4$.  By splitting radial and angular
variables, it can be shown that the eigenmodes decompose as
\begin{equation}\label{eq:VariablesSeparation}
    {\cal Y}_{\beta\ell m}(r,\theta,\phi) = R_{\beta\ell}(r)\,Y_{\ell
m}(\theta,\phi).
\end{equation}
$Y_{\ell m}(\theta,\phi)$ are the spherical harmonics,
related to associated Legendre polynomials $P{}_{\ell}^m$ by
\begin{equation}\label{eq:SphericalHarmonic}
    Y_{\ell m}(\theta,\phi) \equiv \left[ \frac{(2\ell+1)(\ell-m)!}{4
\pi(\ell+m)!}
    \right]^{1/2}\,P{}_{\ell}^m(\cos \theta)\,e^{im\phi}
\end{equation}
and satisfy the relation
\begin{equation}\label{eq:SphericalHarmonic2}
    Y_{\ell m}(\theta,\phi) = (-1)^m Y_{\ell m}^{*}(\theta,\phi).
\end{equation}
The radial eigenfunctions $R_{\beta\ell}(r)$ are solutions of the
radial harmonic equation
\begin{equation}\label{eq:RadialEquation}
    \frac{\sigma^3(r)}{r^2}\,\frac{\d}{\d r}
    \left(\frac{r^2}{\sigma(r)}\,\frac{\d R_{\beta\ell}}{\d r} \right) +
\left[
    q^2 - \sigma^2(r)\frac{\ell(\ell+1)}{r^2} \right]\,R_{\beta\ell} = 0.
\end{equation}
\vskip0.5cm

In flat space ($K=0$) the radial eigenfunctions are simply the
spherical Bessel functions $R_{\beta\ell}=j_{\ell}(\beta r)$.

The cases $K=\pm 1$ can be treated simultaneously by using the
variable
$\xi$ defined by
\begin{equation}\label{eq:ChangeOfVariable}
    \sin \xi=K^{1/2} \frac{r}{\sigma(r)}.
\end{equation}
Note that $\xi$ is related to the dimensionless comoving radial
distance
in units of the curvature radius by $\xi = K^{1/2}\,\chi$. In terms of
these variables the radial equation (\ref{eq:RadialEquation}) takes
the form
\begin{equation}\label{eq:RadialEquation2}
    \frac{1}{\sin^2 \xi}\,\frac{\d}{\d \xi}\left( \sin^2 \xi\,
    \frac{\d R_{\beta\ell}}{\d \xi} \right) + \left[ K q^2 -
    \frac{\ell(\ell+1)}{\sin^2 \xi} \right]\, R_{\beta\ell} = 0.
\end{equation}
Introducing the function $\Pi_{\beta\ell}(\xi)=
R_{\beta\ell}(\xi)\sin^{1/2}(\xi)$ allows to solve the radial equation
in terms of associated Legendre functions $P{}_{\nu}^{\mu}(\cos
\xi)$.

For $K = +1$ (i.e. $\beta^2 = q^2 + 1$)\footnote{We recall that
$\beta^2 = q^2 + K$.}, the radial eigenfunctions are given by
\begin{equation}
    R_{\beta\ell}(\chi) \propto \frac{1}{\sin^2 \chi} P{}_{-1/2 +
    \beta}^{-1/2 - \ell}(\cos\chi)
    \label{eq:SphericEigenFunctions}
\end{equation}
with $\beta \geq{\rm max}(3,\ell+1)$ being an integer\footnote{It is well
known that homogeneous harmonic polynomials of degree $k$ on
$\mathbf{R}^{4}$ restricted to $S^3$ are eigenmodes of the Laplacian
with eigenvalues $k(k+2)$.  It follows that $\beta=k-1$ is necessarily
an integer.}.

For $K = -1$ (i.e. $\beta^2 = q^2 - 1$) the radial eigenfunctions
are given by
\begin{equation}
    R_{\beta \ell}(\chi) \propto \frac{1}{\sinh^2 \chi} P{}_{-1/2 +
    i \beta}^{-1/2 - \ell}(\cosh \chi)
    \label{eq:HyperbolicEigenFunctions}
\end{equation}
and $\beta$ can now take on any positive real value since there are no
periodic boundary conditions to satisfy.

We use the normalisation condition
\begin{equation}\label{eq:Normalization2}
   \int {\cal Y}_{\beta \ell m}^{*} {\cal Y}_{\beta' \ell' m'}
   \frac{r^2 \d r \d \Omega}{\sigma^{3}} = \delta(\beta -
   \beta')\,\delta_{\ell\ell'}\,\delta_{mm'},
\end{equation}
so that the properly normalized functions take the following form
\begin{equation}\label{eq:GeneralEigenFunctions}
    R_{\beta \ell}(\chi) = \left \{
    \begin{array}{ll}
    \displaystyle{ \left( \frac{N_{\beta \ell}}{\sinh \chi}
    \right)^{1/2}\, P{}_{-1/2 + i \beta}^{-1/2 - \ell}(\cosh \chi) } &
    \,K = -1 \\

    \left({2\beta^2}/\pi\right)^{1/2}j_{\ell}(\beta \chi) & \,K = 0 \\

    \displaystyle{ \left( \frac{M_{\beta \ell}}{\sin \chi}
    \right)^{1/2}\, P{}_{-1/2 + \beta}^{-1/2 - \ell}(\cos \chi) } &
    \,K = +1
    \end{array}
    \right.
\end{equation}
with the two coefficients
\begin{equation}\label{eq:Definitions}
    N_{\beta \ell} \equiv \prod_{n=0}^\ell(\beta^2 + n^2) \quad\quad
    M_{\beta \ell} \equiv \prod_{n=0}^\ell(\beta^2 - n^2).
\end{equation}
The radial eigenfunctions (\ref{eq:GeneralEigenFunctions}) differ
from those determined by  Abbott and Schaeffer~\cite{abott86} by an
overall factor $(2\beta^2/\pi)^{-1/2}$ due to the fact that they
used the normalisation
\begin{equation}\label{eq:Normalization1}
\int {\cal Y}_{\beta \ell m}^{*} {\cal Y}_{\beta' \ell' m'}
\frac{r^2 \d r \d \Omega}{\sigma^{3}(r)} = \frac{\pi}{2 \beta^{2}}
\delta(\beta - \beta')\,\delta_{\ell\ell'}\,\delta_{mm'}.
\end{equation}
To finish, in the case of spherical space, the harmonic functions
can be expressed in terms of the 4-dimensional coordinates
(\ref{klein}) as
\begin{eqnarray}\label{A14}
{\cal Y}_{k\ell m}(x)&=&\left[
                \frac{(2\ell+1)(\ell-m)!}{4\pi(\ell+m)!}\right]^{1/2}
                \left(\frac{M_{\beta \ell}}{\sqrt{1-x_0^2}}
\right)^{1/2}
                \nonumber\\
                &&P{}_{-1/2 + \beta}^{-1/2 - \ell}(x_0)
                P{}_{\ell}^m\left(\frac{x_2}{\sqrt{1-x_0^2}}\right)
                \frac{x_1+ix_2}{\sqrt{x_1^2+x_2^2}}.
\end{eqnarray}
\vskip0.5cm

Those expressions are of little value for numerical computation.
There are two routes to compute numerically the eigenmodes.
First, and as explained in Abbott and Schaeffer~\cite{abott86}, one
can use a recursive relation between $R_{\beta \ell}$, $R_{\beta,
\ell-1}$ and $R_{\beta, \ell-2}$. Another efficient
method~\cite{Kosowsky98} makes use of a WKB approximation. These
two methods are complementary.

\ack{ We thank S. Helgason for discussions on the Laplacian operator
during the Williamstown meeting of the American Mathematical Society,
and Alain Riazuelo and Simon Prunet for discussions on the numerical
CMB computations. JW thanks the MacArthur Foundation for its
support. EG thanks FAPESP-Brazil (Proc 01/10328-6) for financial
support.}  \vskip0.5cm

\end{document}